\documentclass[twocolumn]{aastex61}

\usepackage{apjfonts}
\usepackage{graphicx}
\usepackage{amsmath}
\usepackage{amssymb}
\usepackage{color}
\renewcommand{\dh}{\fontencoding{T1}\selectfont{\symbol{240}}} 

\gdef\xx[#1]{\textcolor{red}{#1}}

\newcommand{\thisstar}{KIC\,8462852}	
\newcommand{\kepler}{{\it Kepler}}	
\newcommand{\elsie}{{\it Elsie}}	

\newcommand{\kms}{km~s$^{-1}$}			

\newcommand{\rp}{$r^{\prime}$}
\newcommand{\ip}{$i^{\prime}$}

\newcommand{\caii}{Ca~{\sc ii}}

\newcommand{\nai}{Na~{\sc i}}
\newcommand{\nad}{Na~{\sc i}~D}
\newcommand{\ki}{K~{\sc i}}

\usepackage{natbib}                
\begin{document}

\title[Where's the flux encore]
{The First Post-Kepler Brightness Dips of KIC 8462852} 

\correspondingauthor{Tabetha Boyajian}
\email{boyajian@lsu.edu}

\author[0000-0001-9879-9313]{Tabetha. S. Boyajian}
\affiliation{Department of Physics and Astronomy, Louisiana State University, Baton Rouge, LA  70803 USA}

\author{Roi Alonso}
\affiliation{Instituto de Astrof\' isica de Canarias, 38205 La Laguna, Tenerife, Spain}
\affiliation{Departamento de Astrof\' isica, Universidad de La Laguna, 38206 La Laguna, Tenerife, Spain}

\author{Alex Ammerman}
\affiliation{American Association of Variable Star Observers}
\affiliation{University of Notre Dame QuarkNet Center}

\author{David Armstrong}
\affiliation{Department of Physics, University of Warwick, Gibbet Hill Road, Coventry, CV4 7AL, UK}
\affiliation{Centre for Exoplanets and Habitability, University of Warwick, Gibbet Hill Road, Coventry, CV4 7AL, UK}

\author{A. Asensio Ramos}
\affiliation{Instituto de Astrof\' isica de Canarias, 38205 La Laguna, Tenerife, Spain}
\affiliation{Departamento de Astrof\' isica, Universidad de La Laguna, 38206 La Laguna, Tenerife, Spain}

\author{K. Barkaoui}
\affiliation{Space sciences, Technologies and Astrophysics Research (STAR) Institute, Universit\'e de Li\`ege, Belgium}
\affiliation{Laboratoire LPHEA, Oukaimeden Observatory, Cadi Ayyad University/FSSM, BP 2390, Marrakesh, Morocco}

\author{Thomas G. Beatty}
\affiliation{Department of Astronomy \& Astrophysics, The Pennsylvania State University, 525 Davey Lab, University Park, PA 16802, USA}
\affiliation{Center for Exoplanets and Habitable Worlds, The Pennsylvania State University, 525 Davey Lab, University Park, PA 16802, USA}

\author{Z. Benkhaldoun}
\affiliation{Laboratoire LPHEA, Oukaimeden Observatory, Cadi Ayyad University/FSSM, BP 2390, Marrakesh, Morocco}

\author{Paul Benni}
\affiliation{Acton Sky Portal, Acton, MA 01720, USA}
\affiliation{American Association of Variable Star Observers}

\author{Rory Bentley}
\affiliation{Department of Physics and Astronomy, Louisiana State University, Baton Rouge, LA  70803 USA}

% ...

\begin{abstract}

We present a photometric detection of the first brightness dips of the unique variable star \thisstar\ since the end of the Kepler space mission in 2013 May. Our regular photometric surveillance started in October 2015, and a sequence of dipping began in 2017 May continuing on through the end of 2017, when the star was no longer visible from Earth. We distinguish four main 1--2.5\% dips, named {\it ``Elsie,'' ``Celeste,'' ``Skara Brae,''} and {\it ``Angkor''}, which persist on timescales from several days to weeks. Our main results so far are: (i) there are no apparent changes of the stellar spectrum or polarization during the dips; (ii) the multiband photometry of the dips shows differential reddening favoring non-grey extinction. Therefore, our data are inconsistent with dip models that invoke optically thick material, but rather they are in-line with predictions for an occulter consisting primarily of ordinary dust, where much of the material must be optically thin with a size scale $\ll1 \mu$m, and may also be consistent with models invoking variations intrinsic to the stellar photosphere. Notably, our data do not place constraints on the color of the longer-term ``secular'' dimming, which may be caused by independent processes, or probe different regimes of a single process.

\end{abstract}


\section{Introduction\label{sec:introduction}}

The Planet Hunters citizen science project announced the serendipitous discovery of \thisstar, a peculiar variable star observed by the NASA \kepler\ mission \citep{bor10} from 2009 to 2013 \citep{boy16}. \thisstar's variability manifests itself as asymmetric drops in brightness of up to 22\%, many of which last several days (the ``dips''). There is little or no sign of periodicity in the four years of \kepler\ observations \citep[but see][]{kie17}. Additionally, the duty cycle of the dips is low, occurring for less than 5\% of the four-year period \kepler\ observed it.  Subsequent ground-based follow-up observations to better characterize the star revealed nothing other than \thisstar\ being an ordinary, main-sequence F3 star: no peculiar spectral lines, Doppler shifts indicative of orbiting companions, or signs of youth such as an infrared excess \citep{lis15, mar15, boy16, tho16}.

\explain{This paragraph was slightly re-worded for clarity}
In addition to the short-term variability seen in the \kepler\ long-cadence data, \citet{sch16} discovered a variable secular decline averaging $0.164 \pm 0.013$~magnitudes per century in archival data taken from 1890 to 1989.  While \citet{hip16} claim the dimming found by \citet{sch16} is spurious\footnote{Claims to the contrary by \citet{hip16} are refuted as being due to multiple technical errors:  \url{https://www.centauri-dreams.org/?p=35666}}, the unique longer-term variability was again identified with the \kepler\ full-frame images, which show that \thisstar\ underwent secular dimming by a total of 3\% during the four-year (2009--2013) \kepler\ time baseline \citep{mon16}. More recently, \citet{men17} present over 15 months of space- and ground-based photometry from {\it Swift}, {\it Spitzer}, and {\it AstroLAB IRIS}, showing this variability continues even today.  Further results from ground-based data over 27 months with {\it ASAS-SN} data, and from 2006 to 2017 with {\it ASAS} data also confirm such dimmings, with possible signs of periodic behavior \citep{sim17}.  Thus, \thisstar\ is known to display complex dip-like variations with a continuum of duration timescales ranging from a day to a week to a month to a year to a decade to a century.

\citet{WrightSig2016} re-evaluated the landscape of families of possible solutions that would be consistent with not only the complex dipping patterns, but also the long-term secular dimming. These included broad categories of solutions such as those invoking occulting material in the Solar System, material in the interstellar medium or orbiting an intervening compact object, circumstellar material, and variations intrinsic to the star.  Specific models for the brightness variations have been explored by \citet{kat17}, who modeled a circumstellar ring; \citet{mak16}, who suggested interstellar ``comets''; giant circumstellar exocomets, suggested by \citet{boy16} and modeled by \citet{bod16}; \citet{nes17}, who modeled dust clouds associated with a smaller number of more massive bodies; \citet{bal17}, who suggested a ringed planet and associated Trojan asteroid swarms; \citet{she16}, who find the statistics of the dips to be consistent with intrinsic processes; \citet{met17}, who model the consumption of a secondary body; and \citet{fou17}, who invoke intrinsic variations perhaps related to magnetocovection.

Recently, \citet{wya17} showed that both the dips and dimming could be compatible with circumstellar material distributed unevenly along an elliptical orbit.  These results advocate for additional tests with the various proposed circumstellar scenarios in order to resolve whether the dynamical history of such material could produce the shapes of the dips.  

After the end of the \kepler\ prime mission, we initiated a ground-based monitoring program in order to catch a dimming event in nearly real time.  This paper focuses on the first dip in the {\it Elsie} family, and future papers will focus on the dips that follow.
In Section~\ref{sec:observations}, we describe the observations of the first ground-based detection of a dimming event in \thisstar.  We present several first results from the photometric, spectroscopic, and spectropolarmetric analysis in Sections~\ref{sec:analysis} and \ref{sec:discussion}, and conclude by describing future work.    

\section{Observations}\label{sec:observations}

\subsection{Time-Series Photometry}\label{sec:phot}

\thisstar\ is a northern hemisphere target ($\delta = +44^\circ$) with $V=11.7$\,mag.  From May to September, the star is typically available above airmass 2 for $\sim 4$ to 8\,hr at northern hemisphere observatories, with a decreasing window of visibility until the end of December. In this paper, we present photometric data from several observatories, as described briefly here and in the Appendix.    

Regular photometric monitoring in multiple filters of \thisstar\ started in 2016 March with the Las Cumbres Observatory (LCOGT) 0.4~m telescope network, which consists of telescopes at two northern hemisphere sites: TFN (Canary Islands, Spain) and OGG (Hawaii, USA)\footnote{In 2017 November, an additional northern hemisphere site, ELP (Texas, USA), was added to the LCOGT network.}.  On JD 2,457,892 (UT 2017 May 18; UT dates are used throughout this paper), a drop in brightness was claimed as significant from both TFN and OGG measurements. Observations acquired at Fairborn Observatory corroborated this drop in brightness, and an alert for triggered observations was immediately executed. 

\begin{figure*}[t]
  \begin{center}
          \includegraphics[width=168mm] 
          {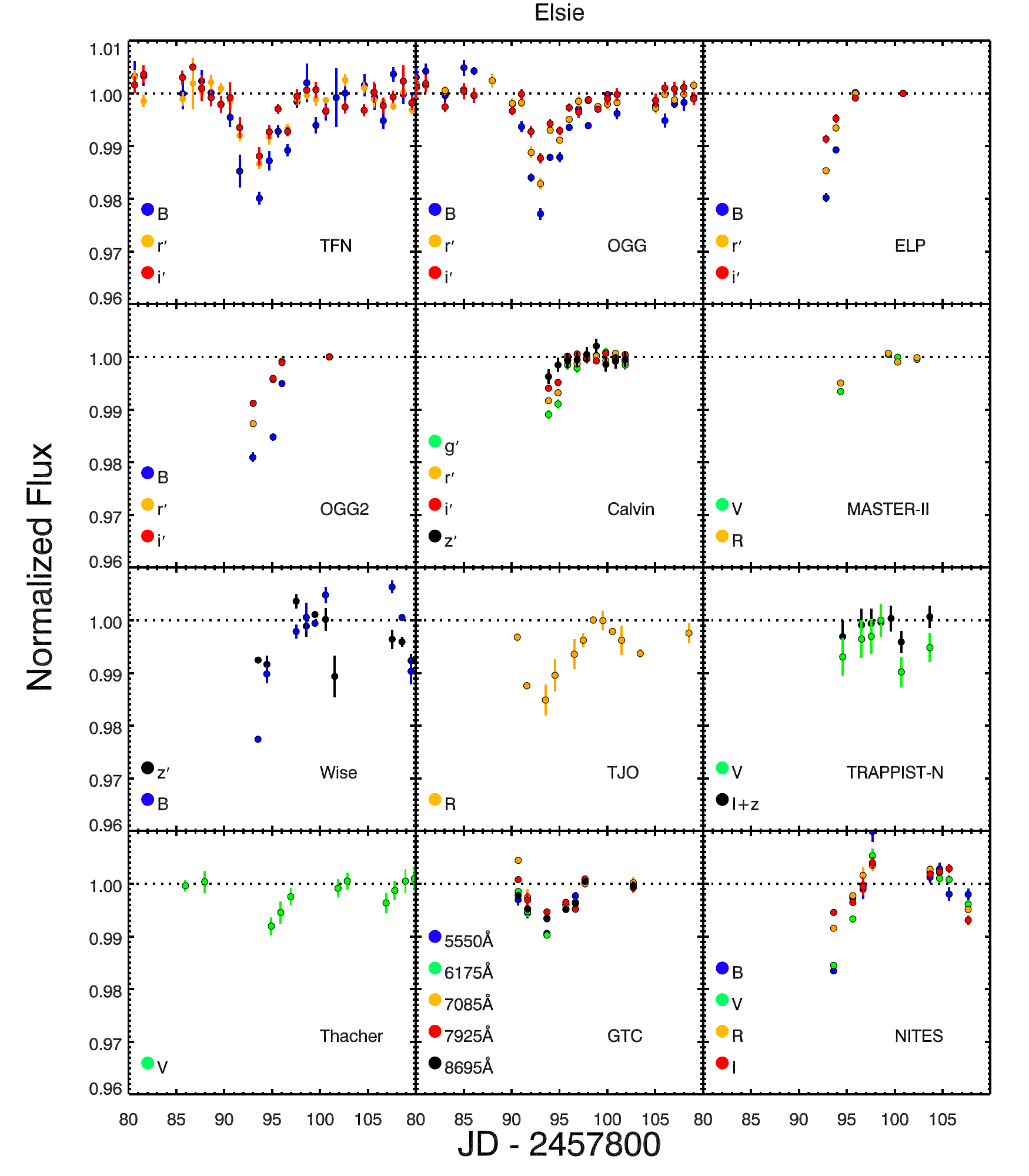}
    \caption{Time-series photometry of {\it Elsie}. See legends for observatory and filter information, and Section~\ref{sec:phot} for details.}\label{fig:phot_montage}
  \end{center}
\end{figure*}

In response to the alert, we acquired additional photometric observations from the Calvin, Master-II, Wise, Joan Or\'o (TJO), Trappist-N, Thatcher, NITES, and Gran Telescopio Canarias (GTC) telescopes.
We showcase this first event observed in 2017 May ({\it ``Elsie''}; Section~\ref{sec:elsie_family}) observed by each observatory in Figure~\ref{fig:phot_montage}, each point of the time series being a nightly average of the differential magnitude in the specified band.
Since most observatories did not have monitoring before the event, we normalize each data set assuming that the stellar flux was at the ``true'' stellar level just after the first event (JD~2,457,900) to ensure consistency.  Note that the analysis carried out in Section~\ref{sec:depths} to determine colors of the dip uses the LCOGT data sets.  These data are normalized using before dip data where the longer baseline improves the normalization (details in Section~\ref{sec:depths}).

\subsubsection{The {\it Elsie} Dip Family}\label{sec:elsie_family} 

\begin{figure*}
 \begin{center}
          \includegraphics[width=\textwidth]
          {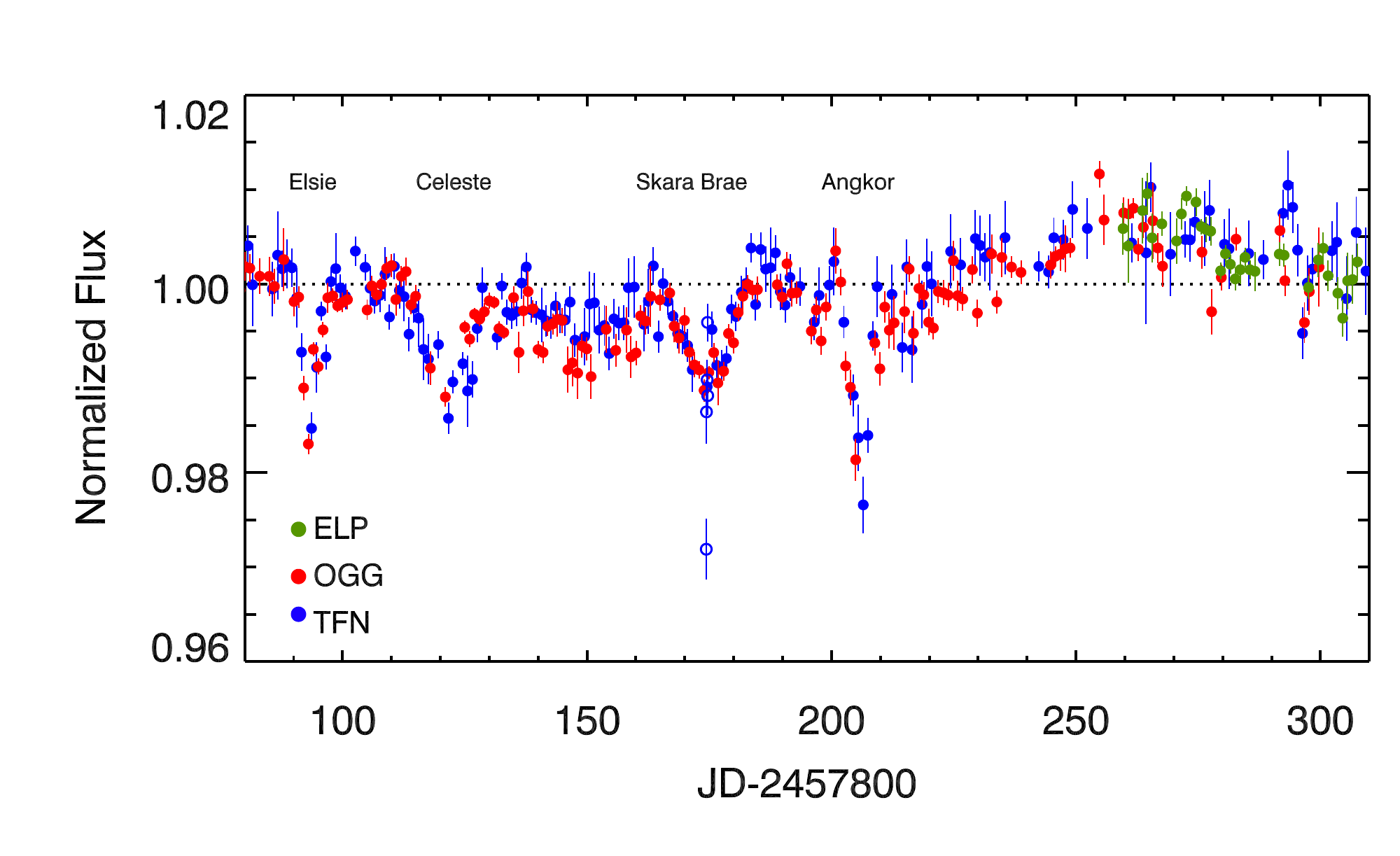}
    \caption{LCOGT time-series photometry of \thisstar\ in the $r^{\prime}$ band from 2017 May through December showing the {\it Elsie} dip family. Each point is a daily average from the LCOGT 0.4~m stations indicated in the legend.  Near the midpoint of {\it Skara Brae}, short-term variability seen over a few hours is indicated as open blue points (see also Figure~\ref{fig:popjupitercorn}). For details, see Section~\ref{sec:phot} and \ref{sec:elsie_family}.}\label{fig:phot_all}
  \end{center}
\end{figure*}

In Figure~\ref{fig:phot_all}, we show the full LCOGT 0.4~m time series from 2017 May through December in the $r^{\prime}$ band. The LCOGT observations are a product of a successful crowdfunding effort through Kickstarter\footnote{\url{https://www.kickstarter.com/projects/608159144/the-most-mysterious-star-in-the-galaxy}}.  As a part of the campaign's design, those who supported the project were eligible to nominate and then vote on the name of the dips. Four major dipping events happened over this time period.  The first, which we named {\it ``Elsie,''}\footnote{The name {\it Elsie} is a play on words with ``L + C,'' short for ``light curve,'' and is also a wink and a nod to the ``L''as ``C''umbres Observatory, for making the project happen.} reaches a minimum around UT 2017 May 19 (JD 2,457,893). Its full duration spans the course of $\sim5$~days.  The temporal gradient of the dip ingress is much sharper than the egress, where the recovery appears to hesitate, making an asymmetric profile overall. Hereafter, we refer to the full period of variability observed in 2017 as the {\it Elsie} dip family.   

A few weeks after the end of {\it Elsie}, the second event, {\it ``Celeste,''}\footnote{This dip [first] appeared to have a slow decline with a quick rise, which is close to a mirror image of {\it Elsie}, which had a quick decline with a slow rise.  {\it Elsie} (or ``L C'') in reverse is ``C L'' or ``ciel,'' which means ``sky'' or ``heavenly'' in French.  {\it ``Celeste''} is the original Latin name from which ``ciel'' is derived.} began to make an appearance, reaching its deepest point around UT 2017 June 18 (JD 2,457,922). {\it Celeste} lasted for a couple of weeks, having a slow decline to its minimum, but then a much more rapid post-minimum rise up.  The rapid egress was stunted about two-thirds of the way up, however, and the flux remained $\sim 0.5$\% below normal brightness levels.  This 30-day-long depression after {\it Celeste} was never given a name, since it was unsubstantial compared to the first two events.  We do however identify this region to be a statistically significant detection of a $\sim 0.5$\% decrease in brightness, which was also independently observed at several other observatories (Bodman et al., in prep.).

\begin{figure}
  \begin{center}
          \includegraphics[trim=15mm 0mm 5mm 0mm, clip, width=84mm]
          {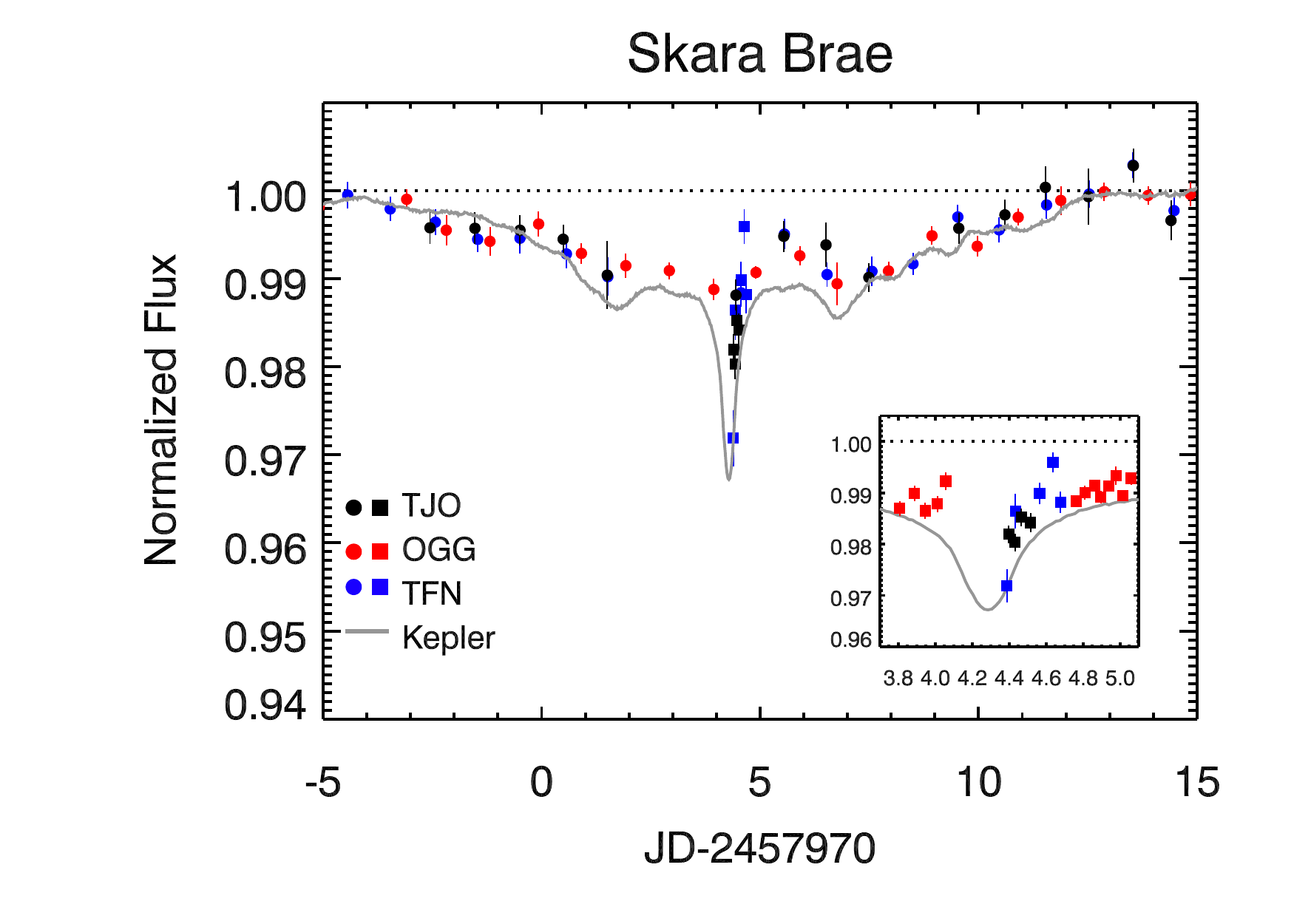}
    \caption{The {\it Skara Brae} event showing daily averages (circles) of ground-based measurements. The solid gray line depicts the \kepler\ long-cadence data from D1540 shifted in time to illustrate the similarity between the events. Near the midpoint of {\it Skara Brae} we show the hourly averages (squares) from LCOGT and TJO to illustrate the short-term variability seen over a span of a few hours.  The plot inset is a close-up view of this midpoint region. See Section~\ref{sec:elsie_family} for details.}\label{fig:popjupitercorn}
  \end{center}
\end{figure}

The third event, {\it ``Skara Brae,''}\footnote{The brightness variations for \thisstar\ share some of the same traits as this lost city located in the far north of Scotland. They're ancient; we are watching things that happened more than 1000\,yr ago. They're almost certainly caused by something ordinary, at least on a cosmic scale. And yet that makes them more interesting, not less. But most of all, they're mysterious. What the heck was going on there, all those centuries ago?} began to appear just when the month-long depression after {\it Celeste} showed possible signs of recovery.  {\it Skara Brae} behaved very similarly to {\it Celeste} in a mirror-image profile, with a $\sim 1$\% depth in $r^{\prime}$-band lasting on the order of two weeks. In the middle of {\it Skara Brae} (on UT 2017 August 8; JD 2,457,974) there was an additional narrow (few hour) $\sim2$\% dip. We did not observe the ingress of this dip (which could have lasted up to 7\,hr) because this region fell within a gap between observatories, but observations from both LCOGT and TJO caught the rapid return to the level of the broad dip over a 4\,hr period. This event was the first and only significant short-timescale ($<1$~d) variability observed in the 2017 {\it Elsie} family of dips (Figure~\ref{fig:phot_all}, open circles). We show a close-up of this feature in Figure~\ref{fig:popjupitercorn}.  While this is a seemingly drastic change in brightness within the \elsie\ family, we find that the dip gradient here ($\sim 0.2$\% hr$^{-1}$, or $\sim 5$\% day$^{-1}$) is not extreme compared to the dip gradients observed by \kepler, which can be up to a factor of 10 steeper \citep{boy16}.

This smaller gradient could indicate that the current dips are less optically thick than those with steeper gradients by \emph{Kepler}, which would be consistent with the smaller dip depths, although other geometric factors may also play a role.  In any case, the \kepler\ observations still provide the most stringent constraint on the orbit of the parent body (or bodies) from consideration of the light-curve gradient \citep{boy16}. We also find the duration of this event is no shorter than the shortest seen by \emph{Kepler}, which was 0.4~d.  Finally, we note that {\it Skara Brae}'s full contour -- a symmetrical broad shallow dip containing a brief deep dip within -- is unique\footnote{``Think of the dip's creator being something like a giant Jupiter-sized kernel of corn, that halfway across the star pops into a {\it popjupitercorn}, for maybe a few hours, then spontaneously de-pops.'' -- T. Hicks; August 2017.}.  This allows us to compare its appearance with the dips observed by \kepler, where only a couple were at a similar level of 1--2\%.  We find the \kepler\ dip at D1540 \citep{boy16} being remarkably similar in depth, shape, and duration (see Figure~\ref{fig:popjupitercorn} for a direct comparison).

{\it ``Angkor,''}\footnote{Following the theme of lost cities, Angkor is perhaps the greatest lost city of medieval times.} the fourth significant dip in the complex, appeared two weeks after {\it Skara Brae}, reaching its deepest depth around UT 2017 September 9 ($\sim$ JD 2,458,006). The {\it Angkor} ingress exhibited the most rapid, sustained ($\sim 4$ days long) flux gradient observed for the star from the ground to date, and the egress was similarly rapid, though it remained at a depth of $\sim 0.5$\% below normal for about a week before brightening to slightly above normal levels.  The post-{\it Angkor} monitoring data rose up $\sim 0.5$\% compared to pre-{\it Elsie} levels (dotted line, Figure~\ref{fig:phot_all}).  Around the end of 2017 October (JD 2,458,050) this brightening trend reversed, returning back to unity by mid-December ($\sim$ JD 2,458,100).   

\subsection{Spectroscopy}\label{sec:spectra} 

Over a six month period beginning 2017 May 20, we acquired low-resolution spectra with the Kast Double Spectrograph mounted on the 3~m
Shane telescope \citep{miller-stone93} at Lick Observatory, as well as from the DEep Imaging Multi-Object Spectrograph
\citep[DEIMOS;][]{fab03} and the Low Resolution Imaging
Spectrometer using the Keck telescopes.
All of the Lick and Keck low-resolution
spectra are broadly consistent with the object being a
main-sequence F star; a future paper will present them
in greater detail (Mart\'inez Gonz\'alez, in prep).

A total of eight high-resolution spectra were acquired with the HIRES spectrograph on Keck~I.  Five of the eight were taken prior (circa 2015, 2016), and the other three were taken during the {\it Elsie} event.
First, we checked for radial velocity (RV) variations. 
We fine-tuned the wavelength solution using 
the numerous telluric lines 
in the vicinity of \nad\ using the telluric spectrum from the \citet{Wallace2011} solar atlas,
applied barycentric velocity corrections calculated with the IDL code
\texttt{BARYCORR} \citep{barycenter}, 
verified that the interstellar medium (ISM) lines remained fixed, 
and measured RV offsets between 
each out-of-dip spectrum along with their coadded 
average and the three in-dip spectra, 
resulting in $\Delta$RV $= 0.38 \pm 0.35$~\kms.
We are unable to discern RV variability with the present quality of the wavelength solutions during these epochs, limiting the presence of anything larger than a gas giant within 0.1~au.

\begin{figure*}
\begin{center}
\plottwo{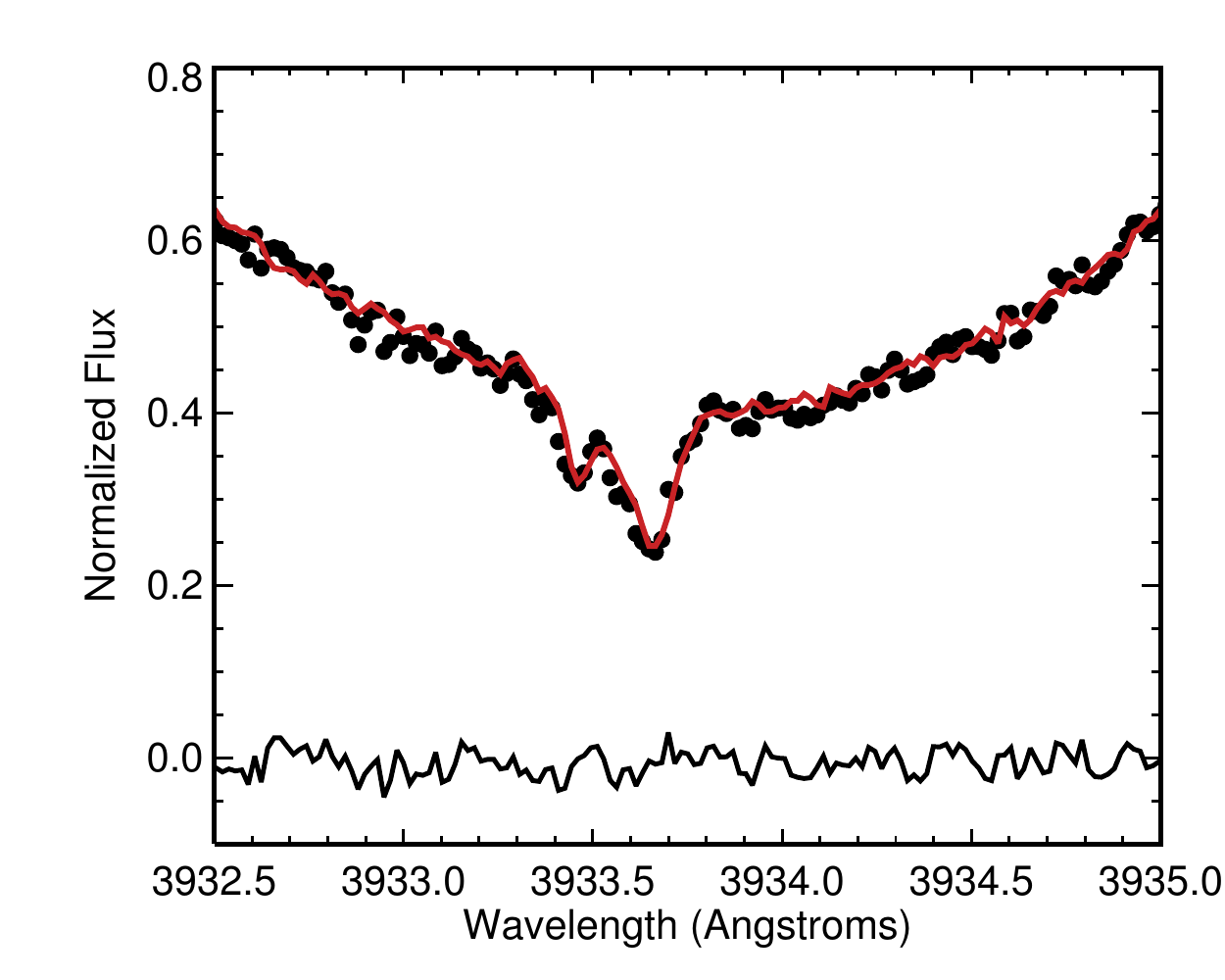}{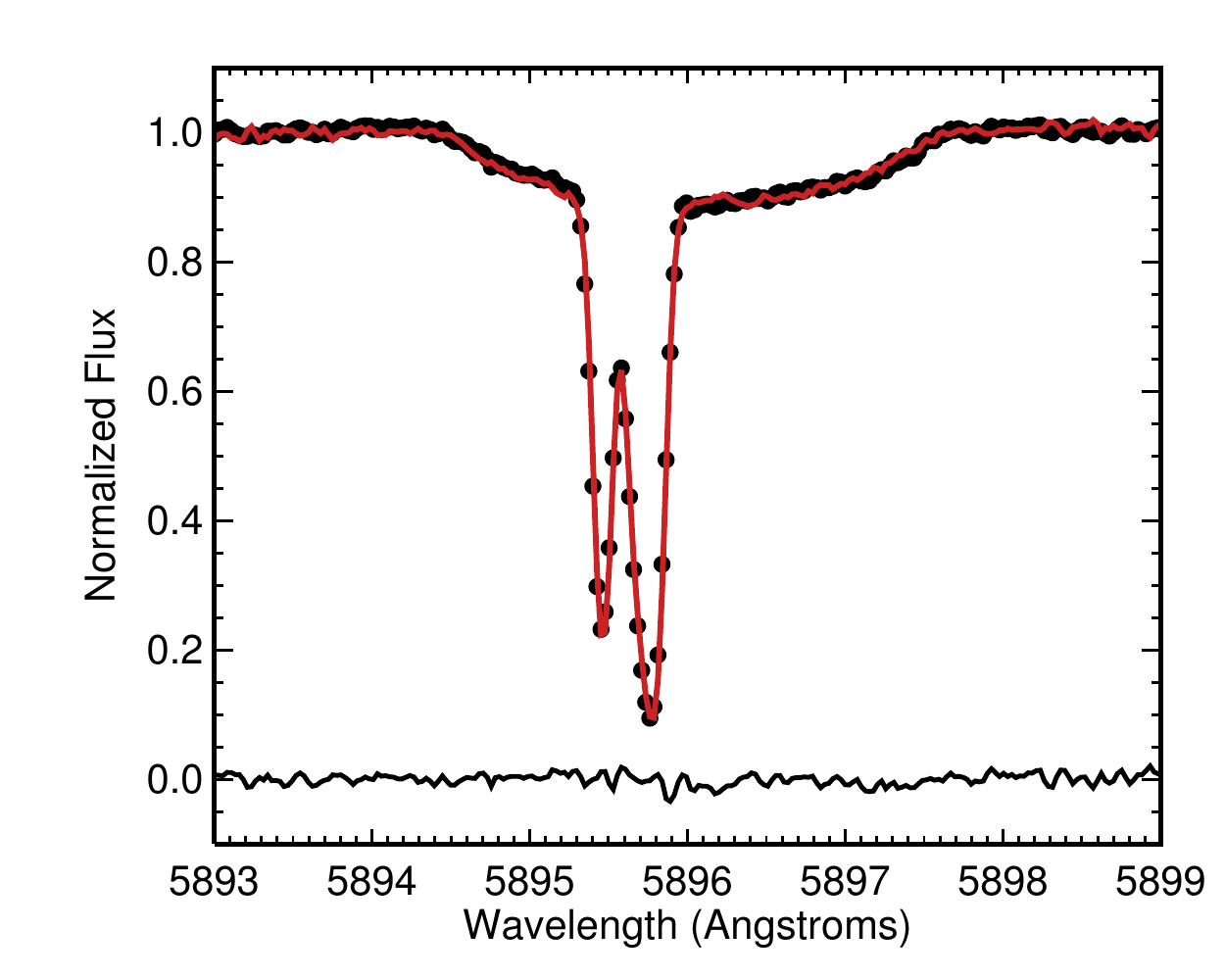}
    \caption{HIRES spectra of \thisstar\ at the
    \caii~K (left) and \nad1 (right) lines
    prior to (black dots) and during (red line) 
    the {\it Elsie} event, 
    along with residuals (prior minus during). Error bars are smaller than the black symbols. We measure no significant variation in stellar RV ($\Delta < 0.4$~\kms) or equivalent width of the ISM lines ($<$1\%), as evidenced by the residuals. See Section~\ref{sec:spectra} for details.}\label{f:ISM}
  \end{center}
\end{figure*}

The ISM imprints absorption lines in spectra at 
\caii\ (H\&K at 3968.469~\AA\ and 3933.663~\AA),
\nai\ (D2 at 5889.951~\AA, D1 at 5895.924~\AA), and
\ki\ (7698.964~\AA);
unfortunately, our in-dip spectra do not reach \ki.
Close-up views of the \caii~K and \nai~D1 spectral regions  
are shown in Figure~\ref{f:ISM}, along with their residuals (prior minus during \elsie). 
The foreground ISM can be described by three Gaussians 
(the profile of the redder of the two apparent components is 
asymmetric and is clearly blended with at least one additional component).
We find no significant variations in the depths of these ISM lines 
(i.e., differences in equivalent widths of the ISM lines are $<$1\%),
following the procedures outlined 
by \citet{WrightSig2016} and \citet{Curtis2017}. 
The amplitude of the residuals at these ISM lines is on par with variability due to the S/N, 
and it will be challenging to discern real changes in interstellar absorption at or below this level 
during future events without higher-quality spectra (or larger dips).

\subsection{Infrared Photometry}\label{sec:neowise}

\begin{figure}
  \begin{center}
          \includegraphics[trim=10mm 5mm 0mm 10mm, clip, width=84mm]
          {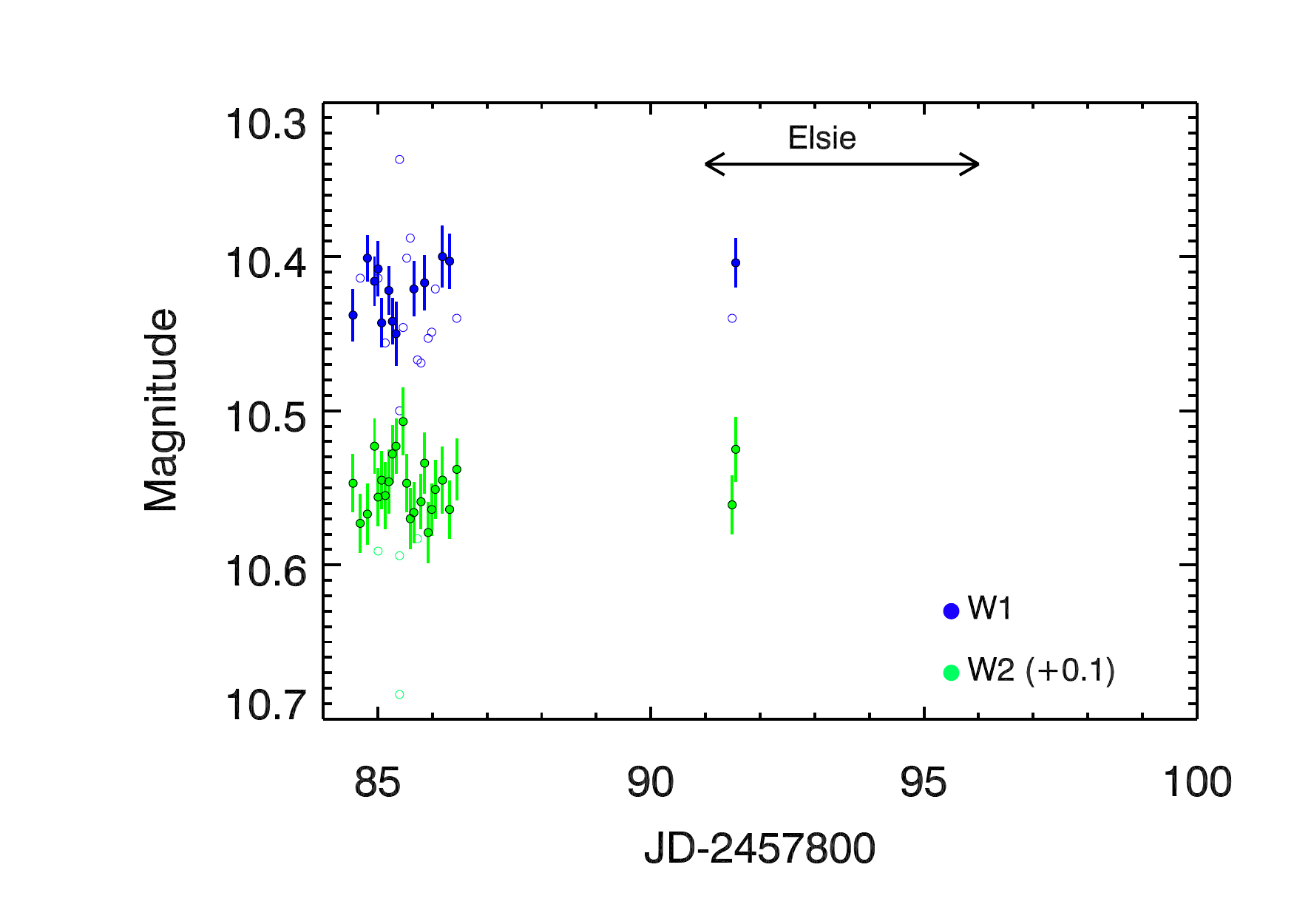}
    \caption{NEOWISE time series photometry of \thisstar\ in the W1 (3.4\,$\mu$m) and W2 (4.6\,$\mu$m) bands according to the legend at the bottom right. Measurements with PSF-fitting quality metric $\chi^2_{\rm PSF} < 2$ are shown as filled points with their associated 1$\sigma$ uncertainties; otherwise the point is unfilled and without uncertainties for clarity.  The approximate timing of \elsie\ identified in optical data is marked on the graph. See Section~\ref{sec:neowise} for details.}\label{fig:neowise_elsie}
  \end{center}
\end{figure}

NEOWISE observations of \thisstar\ in the W1 (3.4~$\mu$m) and W2 (4.6~$\mu$m) bands were acquired a week before \elsie, and seren{\it dip}itously on  JD 2,457,892 (around the time of the largest optical flux decrease for \elsie) with the spacecraft's occasional toggle to avoid the Moon. All NEOWISE observations of \thisstar\ were extracted without any cuts on flag parameters. Inspecting each detection, we see that all observations had ph\_qual ratings of ``A'' in both bands, and none of the detections was affected by latent images, other persistence features, or diffraction spikes as would be indicated by the ``ccflags'' parameter \citep{cut15}. We show this temporal series in Figure~\ref{fig:neowise_elsie}.  We find no detectable change in brightness with the observations taken during {\it Elsie} compared to the week prior in either of the NEOWISE bands.
Work is currently underway analyzing near-IR measurements (Clemens et al., in prep.) and additional mid-IR measurements (Meng et al., in prep.) taken throughout the \elsie\ family of dips to put additional constraints on variability after this first event.

\subsection{Polarimetry}\label{sec:polarimetry}

Evidence for the scattering of stellar light by circumstellar dust can be probed with polarimetry. Spectropolarimetry of \thisstar\ was obtained with the Kast spectropolarimeter on the 3~m Shane reflector at Lick Observatory (see \citealt{mau14} for a description of the Kast instrument) at eight epochs: 2017 May 20.4; June 2.3, 21.4, 27.4; July 1.4, 16.4, 17.4; and August 1.4.  On each night, the source was observed 18 minutes for each Stokes parameter $Q$ and $U$, for a total of 36 minutes on-source integration time. The polarization $P$ and position angle $\theta$ were derived according to the methodology described by \citet{mau14}. On every night, two strongly polarized standard stars (Hiltner\,960 and HD\,204827) and one weakly polarized star (HD\,212311) were observed for calibration purposes. HD\,204827, in particular, was used to calibrate the position angle on the sky.

\begin{figure*}
\begin{center}
\plottwo{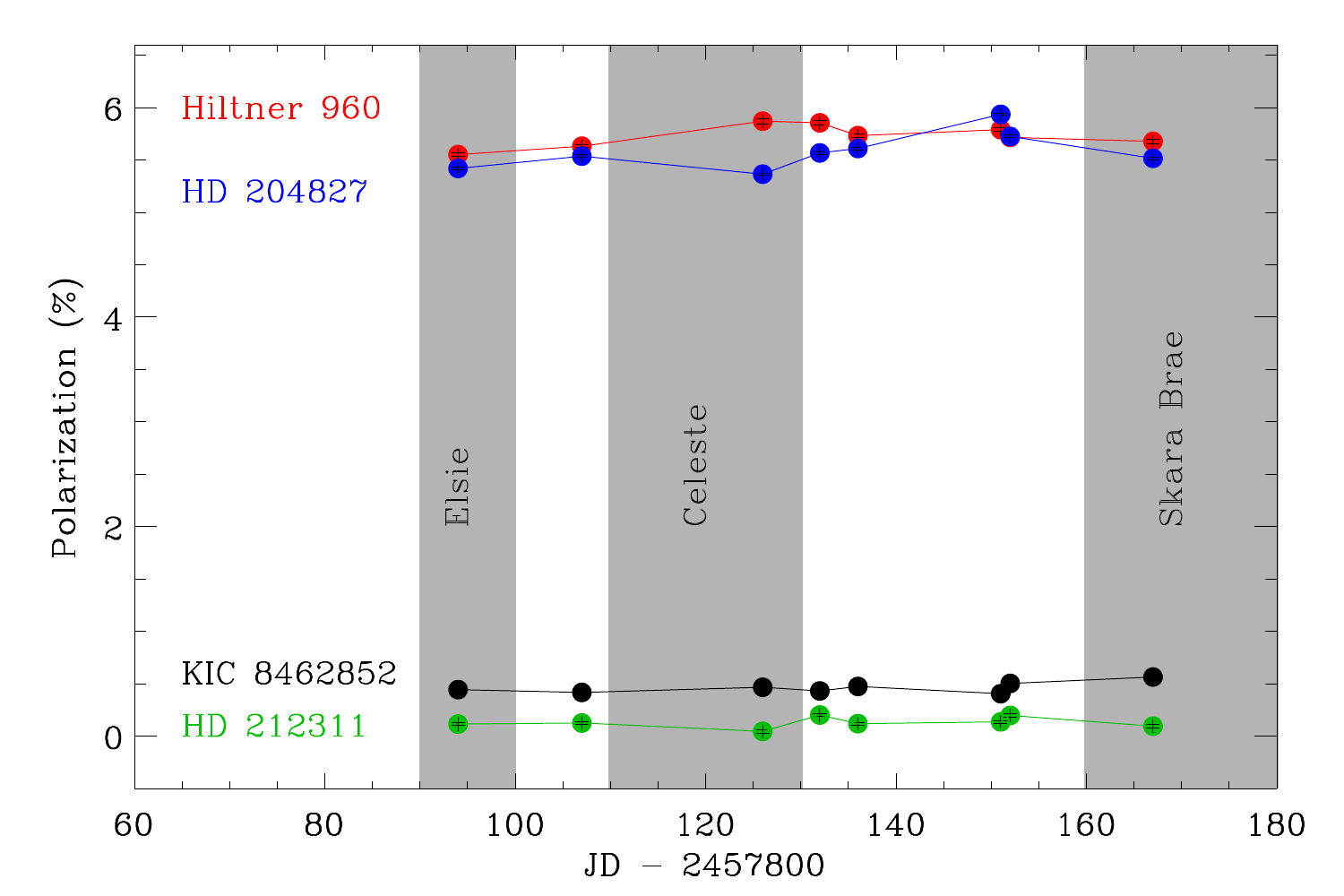}{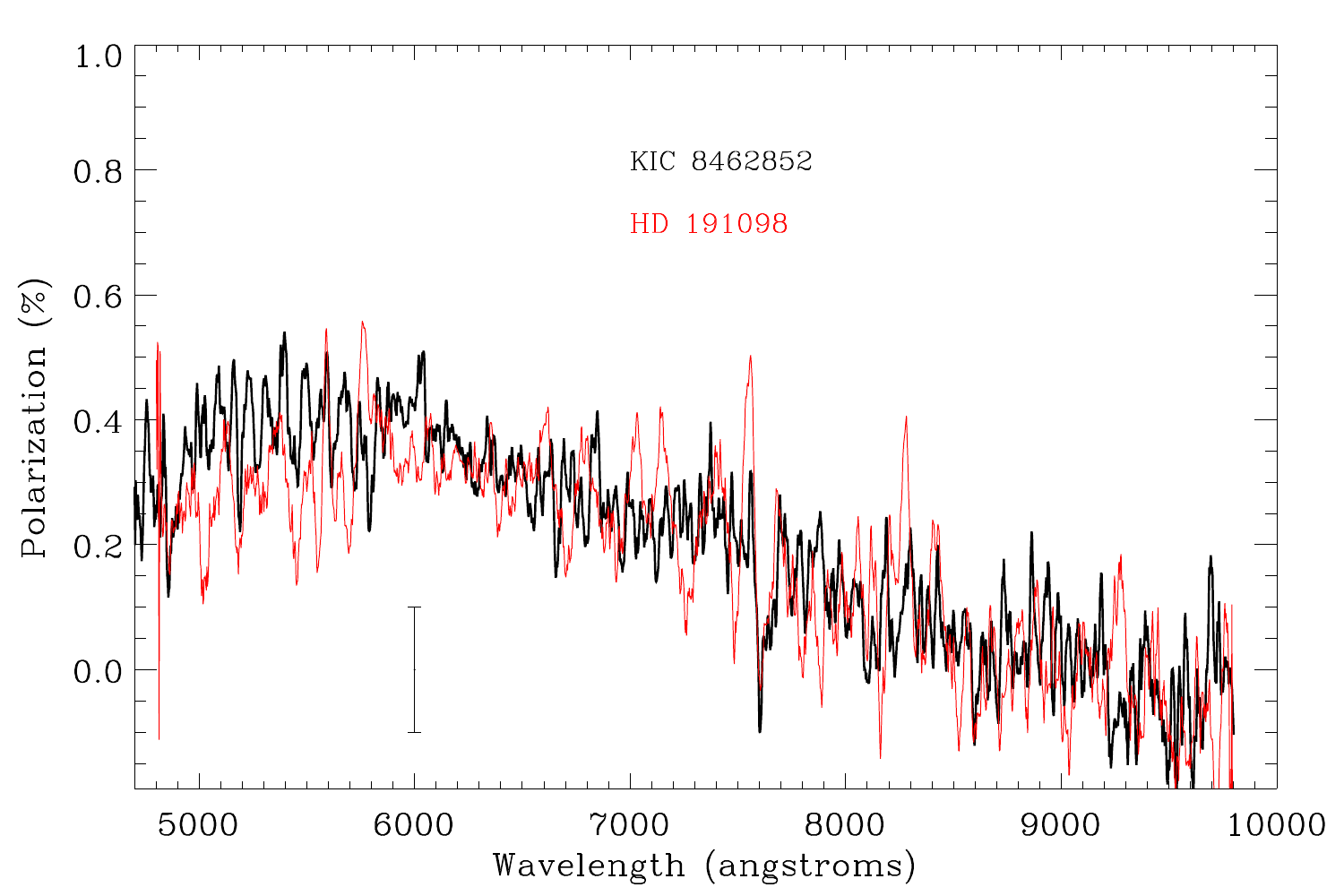}
    \caption{{\it Left:} Polarization over time for \thisstar\ (black), Hiltner~960 (red), HD~204827 (blue), and HD\,212311 (green). The statistical uncertainties appear smaller than the data points.  {\it Right:} Polarization and position-angle spectra of \thisstar\ (black curve). The data are the average of all seven epochs of observations performed in 2017 May--August. The red curve represents data for a nearby star HD~191098, which was used as a probe of interstellar polarization. The noise fluctuations of \thisstar\ reflect the total statistical and systematic measurement uncertainty, which we estimate at $\pm0.10$\% (graphically represented near bottom center).  See Section~\ref{sec:polarimetry} for details.}\label{fig:specpol}
  \end{center}
\end{figure*}

Over the course of our eight epochs KIC~8462852 exhibited an average (standard deviation) polarization of $P=0.46(0.05)$\% and position angle $\theta=91.0^{\circ}(2.6^{\circ})$ in the $V$ band (5050--5950\,\AA). The statistical uncertainty in $P$ was typically 0.01\% on a given epoch, but in our experience with Kast the systematic uncertainties are usually at least 0.1--0.15\%. Over the same period the average (standard deviation) $V$-band polarization of our polarized standard Hiltner\,960 is $P=5.58(0.18)$\%, $\theta=54.9^{\circ}(0.7^{\circ})$, consistent with published values \citep{sch92}. The weakly polarized star HD~212311 exhibited $P=0.13(0.05)$\%, within 2$\sigma$ of the \citet{sch92} value. The results are illustrated in Figure~\ref{fig:specpol}, and indicate that KIC~8462852 did not vary significantly in polarization, relative to our calibration stars. 

The polarization characteristics of KIC~8462852 appear to be consistent with interstellar polarization, presumably induced by the dichroic absorption of light by nonspherical dust grains oriented along the Galactic magnetic field. Figure~\ref{fig:specpol} illustrates the polarization and position-angle spectra of the source, along with the F5~V star HD~191098 ($V=10.09$\,mag; $0.2^{\circ}$ from KIC~8462852 on the sky), which we used as a probe of the approximate interstellar polarization on 2017 June 2; the spectroscopic parallax distance of this star ($M_V=3.5$\,mag) is $\sim201$\,pc, which is not ideal, since KIC~8462852 lies at an much greater estimated distance of 391\,pc. Nonetheless, HD~191098 can provide a useful probing of interstellar polarization if the majority of intervening material is closer to Earth than both sources. The spectropolarimetric data for both KIC~8462852 and HD~191098 follow the wavelength-dependent ``Serkowski'' form \citep{ser75} that is expected for interstellar polarization from Galactic dust (where the peak polarization is typically seen near 5300--5400\,\AA). HD~191098 exhibits a $V$-band polarization of $P\approx0.40$\%, $\theta \approx 78^{\circ}$, only slightly lower than that of KIC~8462852 in magnitude and offset by $\sim14^{\circ}$ in position angle. Moreover, we examined the polarization of three stars from the catalog of \citet{hei00} that lie within $2^\circ$ of KIC~8462852 on the sky: HD~190149, HD~191423, and HD~191546. The average (standard deviation) polarization and position angle exhibited by these stars is $P=0.47(0.36)$\% and 113$^{\circ}$(18$^{\circ}$). This is also comparable to our measurements of KIC~8462852. 

The polarization we measure appears to be significantly lower than the broad-band imaging polarimetry values that \citet{ste18} have reported with the RINGO3 instrument \citep{slo16} over a similar time period. For example, over May through August 2017, they reported a non-variable source ($<0.2$\% fluctuation) with average polarization and position angle of $P=1.2\%(0.2\%)$, $\theta=72^{\circ}(6^{\circ})$ for the $b^{*}$ band (3500--6400\,\AA); $P=0.6\%(0.1\%)$, $\theta=74^{\circ}(6^{\circ})$ for the $g^{*}$ band (6500--7600\,\AA); and $P=0.6\%(0.1\%)$, $\theta=73^{\circ}(6^{\circ})$ for the $r^{*}$ band (7700--10,000\,\AA). However, the level of discrepancy between these measurements and ours is not particularly surprising, given the large systematic uncertainties associated with the RINGO3 instrument \citep{slo16}.  Regardless of systematics, we reiterate that the results here and those from \citet{ste18} both show the object has polarization properties typical of its location on the sky, and limit any change in optical polarization between in and out of dip epochs of $<0.05$\% (Kast) and $<0.1 - 0.2$\% (RINGO3). 
Finally, we note that analysis of near-IR polarimetry taken during the {\it Celeste} event are underway, and should be able to provide additional constraints for the system and its environment (Clemens et al., in prep.).

\subsection{Radio SETI}\label{sec:seti}

The Allen Telescope (SETI Institute) performed a series of \thisstar\ radio observations during the \elsie\ dimming event. These were subsequent to the observations in 2015 \citep{har16}. From 2017 July 8 through 2017 August 8, 1022 separate 92-s observations were performed which resulted in scanning 1419 MHz to 4303 MHz for radio signals. No narrow-band radio signals were found at a level of 180--300\,Jy in a 1\,Hz channel, or medium-band signals above 10\,Jy in a 100\,kHz channel, which would correspond to transmitters at the distance of KIC~8462852 having effective isotropic radiated power (EIRP) of (4--7) $\times 10^{15}$~W and $10^{19}$~W for the narrow-band and moderate-band observations. While orders of magnitude more powerful than the planetary radar transmitter at Arecibo Observatory, $2 \times 10^{13}$~W EIRP, the actual power requirements would be greatly reduced if emissions were beamed directly at Earth.

\section{Analysis}\label{sec:analysis}

\subsection{Size, Shape, and Wavelength Dependence of {\it Elsie}}\label{sec:depths}

In the discussion presented here, we limit the range of our analysis to the first event, \elsie; the analysis of the remaining \elsie\ dip family will be presented by Bodman et al. (in prep.). 
We also assume that the differential fluxes found in each filter are measures of the additional opacity causing the dips, not the total color of all the material along the line of sight (i.e., we apply no correction in the baseline flux for the longer-term brightness variations).  For the following discussion, we define depth ratios, $B/i^{\prime}$ and $r^{\prime}/i^{\prime}$, as the ratio of the dip depths measured in normalized flux in the $B$ to $i^{\prime}$ filters and $r^{\prime}$ to $i^{\prime}$ filters, respectively.

Continuous time-series coverage throughout the \elsie\ event shows the largest depth of $2.5$\% in the $B$-band on JD 2,557,893. With monitoring across the many observatories, we obtained a nearly continuous cadence time series during \elsie, and thus can rule out a missed detection of a deep, short-lived dip, with the longest data gap occurring daily from 15 to 19 UT (a duration of $\sim 0.2$~days). We fit the LCOGT multiband light curve with an $N+8$ parameter model, where $N$ is the number of days over which the fitted data are taken. The unbinned data for each site are used for the fit.

\begin{figure}
\includegraphics[width=3.0in, trim= 0 0 0 0 ]{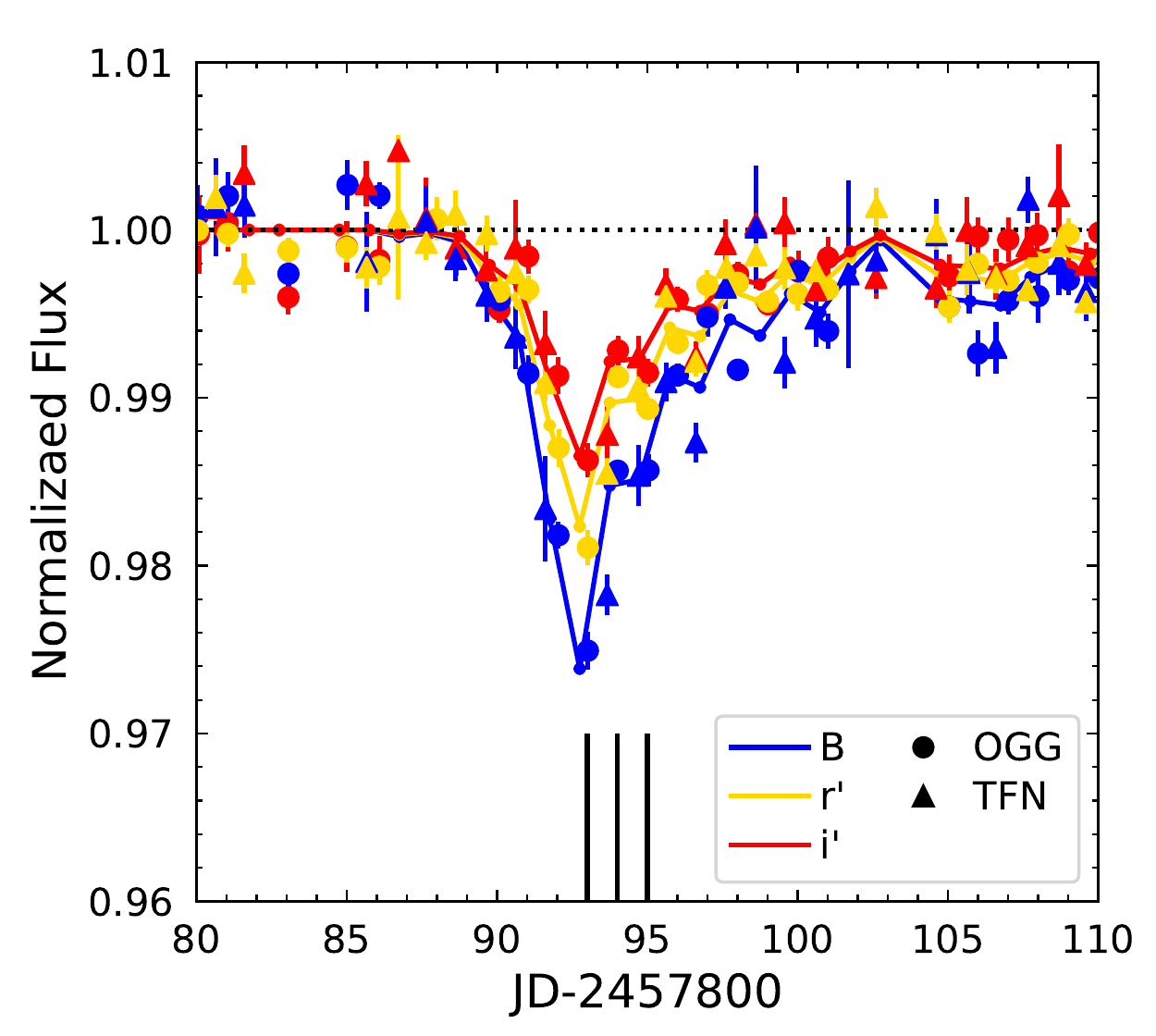}
\caption{Plot of the \elsie\ dip with $N+8$ model fit. Red, yellow, and blue indicate $i^{\prime}$, $r^{\prime}$, and $B$ filters (respectively) for the data points and fit lines. The different observatory sites are indicated by different marker shapes as depicted in the legend, and each data point is the average daily value. Our fit results in the colors of \elsie\ to be $B/i^{\prime} = 1.94 \pm 0.06$ and $r^{\prime}/i^{\prime} = 1.31 \pm 0.04$. See Section~\ref{sec:depths} for details. The black vertical lines indicate when Keck/HIRES spectra were taken (see Section~\ref{sec:spectra}).
}
\label{fig:Bayes}
\end{figure}

A single depth ratio for each filter set ($B/i^{\prime}$ and $r^{\prime}/i^{\prime}$) is assumed for the entire dip. We include normalizations ($n$) of each telescope in each filter as free parameters, resulting in a total of six normalizations, to ensure that the different sites are normalized in a consistent manner. Each day has a depth ($d_i$) in the $i^{\prime}$ filter fit to the data binned in that day. When $d_i$ is fit, it is compared to the $i^{\prime}$ band for each applicable site with normalization, and to the $B$ and $r^{\prime}$ data after multiplied by the depth ratio ($R$) and normalization, e.g., $F_\mathrm{Filter}=n_\mathrm{Filter}(1-d_i\times R_\mathrm{Filter})$. This allows for the color ratios, the normalizations, and the depths to be fit simultaneously.
 In order to prevent the normalizations from wandering too far from 1.0, we include ten days of pre-\elsie\ data where the depths are set to 0. To overcome degeneracies in the fit, we explored the parameter space with the Markov Chain Monte Carlo method using the ``emcee'' package \citep{emcee13}. We used flat priors over ranges 0.5--5, 0.99--1.01, and 0--0.2 for the depth ratios, normalizations, and depths (respectively), with the results of a least-squares fit as the starting point. The results of the Bayesian fit are shown in Figure \ref{fig:Bayes}, and the depth ratios are $B/i^{\prime}=1.94\pm0.06$ and $r^{\prime}/i^{\prime}=1.31\pm0.04$. The $r^{\prime}/i^{\prime}$ ratio measured at Calvin Observatory (1.29) is in agreement with the LCOGT value.  The extinction due to optically thin dust is typically characterized by a higher extinction at bluer wavelengths such as is observed for \thisstar\ during the \elsie\ dip. The colors observed indicate dust grain sizes larger than are seen for interstellar dust and are more likely to arise from circumstellar dust \citep{sav79, men17}.

\subsection{Particle Size and Chemical Composition}\label{sec:comp}

Under the assumption that the dimming is caused by intervening dust, the multicolor observations taken during \elsie\ may be used to put some constraints on the particle size and chemical composition. We therefore compared the measured amount of attenuation in different filters found in the previous Section, $B/i^{\prime}=1.94\pm0.06$ and $r^{\prime}/i^{\prime}=1.31\pm0.04$, with the theoretical ratios of dust opacities for optically thin distributions at those wavelengths. The opacities of grains for different chemical compositions and different particle sizes are taken from the tables of \cite{budaj15}. These opacities assume homogeneous spherical dust grains with a relatively narrow Deirmendjian particle size distribution (FWHM~$\approx 0.4$~dex; \citealt{dei64}), as well as effective wavelengths of $B$, $r^{\prime}$, and $i^{\prime}$ of 0.4361, 0.6215, 0.7545~$\mu$m, respectively. 

We explore the signatures of a few refractory dust species and water ice. Silicates are perhaps the most important refractory species. They are divided into the family of pyroxenes and olivines.  Optical properties of silicates are quite sensitive to the amount of iron in the mineral and that is why we explore iron free and iron rich cases. Namely, for pyroxenes (Mg$_{x}$Fe$_{1-x}$SiO$_{3}$) we consider the two cases: a mineral with zero iron content ($x=1$) which is called enstatite and a pyroxene with 60/40 iron/magnesium ratio. The refractive indexes of these pyroxenes are taken from \citet{Dorschner95}.  Olivines (Mg$_{2y}$Fe$_{2-2y}$SiO$_{4}$) may also have different iron content. Again we consider the mineral with zero iron content ($y=1$) called forsterite, and use its refractive index from \citet{jager03}. We also consider an olivine with 50/50 iron to magnesium ratio, taking its index of refraction from \citet{Dorschner95}. Alumina (Al$_{2}$O$_{3}$) is one of the most refractory species which might be present in such environments and we take the complex refractive index for $\gamma$ alumina from \citet{Koike95}. The refractive index for iron is taken from \citet{johnson74}. Those of carbon are from \citet{jager98}, and they assume carbon at high temperature (1000~$^\circ$C). Lastly, the complex index of refraction for water ice is taken from \citet{warren08}.

\begin{figure*}
\begin{center}
\plottwo{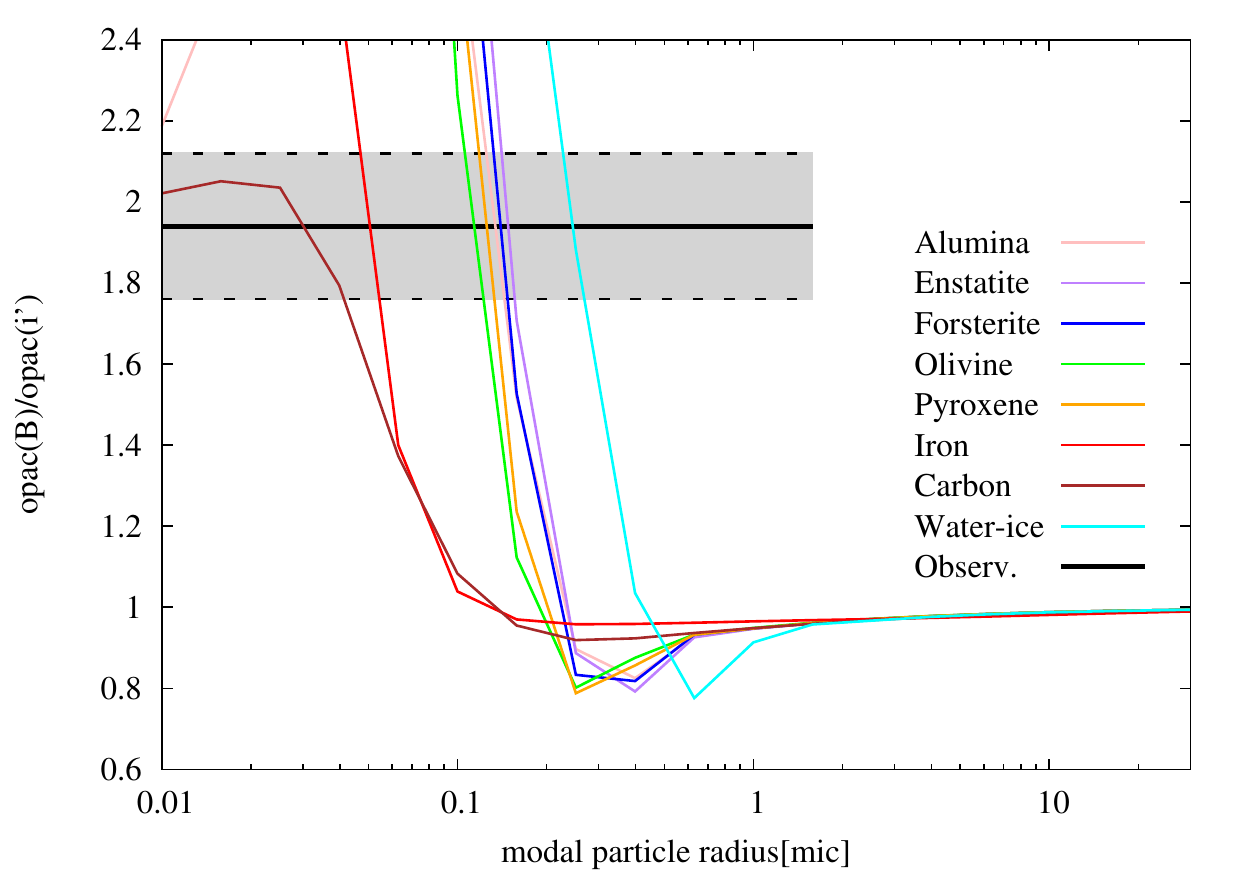}{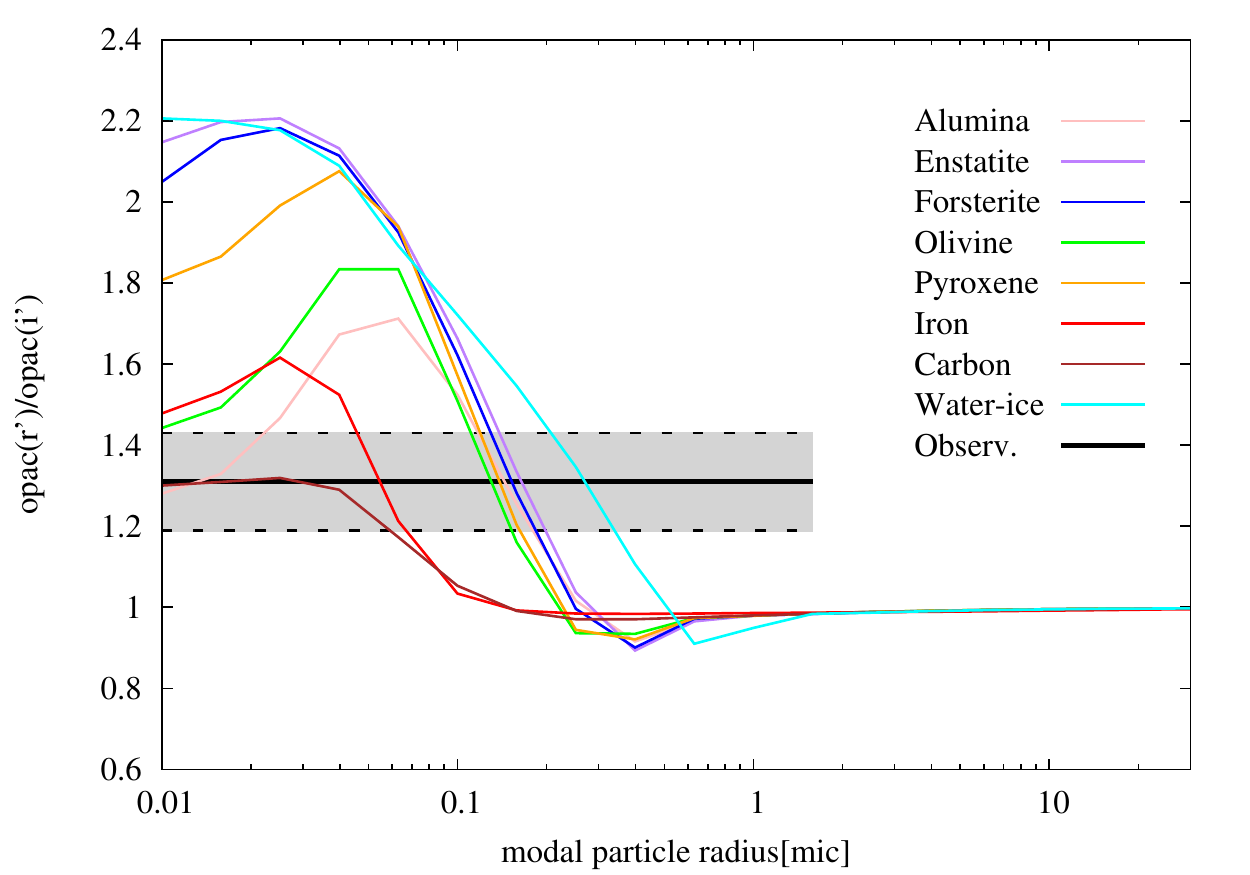}
    \caption{Opacity ratio in $B/i^{\prime}$ (left) and $r^{\prime}/i^{\prime}$ as a function of modal particle radius for several dust species as indicated in the legend. Observations with $3\sigma$ limits are plotted as horizontal lines.  See Section~\ref{sec:comp} for details.}\label{fig:opac}
  \end{center}
\end{figure*}

The comparison between the observed and theoretical opacity ratios for different modal size (i.e., radius) of the grains and their chemical composition is shown in Figure~\ref{fig:opac}. 
We find that most of the silicates and alumina should have particle size of about 0.1--0.2\,$\mu$m. If the dust were composed solely from iron, it would have significantly smaller particles of about 0.04--0.06\,$\mu$m. Carbon would require even smaller particles, $<0.06$\,$\mu$m. On the other hand, water ice would require 0.2--0.3\,$\mu$m particles.  We conclude that regardless of composition, the dust is $\ll 1$\,$\mu$m in size and optically thin, which is consistent with the predictions in Figure~3 of \citet{wya17}.  Circumstellar dust that is this small would be strongly affected by radiation forces, putting it on an unbound orbit as soon as it was created. Thus, if it is circumstellar, the dust causing these short-duration dips must have been created recently, since radiation forces would cause it to spread from its point of origin.

\section{Discussion}\label{sec:discussion}

The detection of dips from the ground extends the duration of dip activity in \thisstar\ from the 4\,yr of the {\it Kepler} mission, to 8\,yr since the beginning of the {\it Kepler} mission.   The complexity and duration of the \elsie\ dip family is reminiscent of the Quarter 16 (Q16) dip complex during the \kepler\ mission (see Figure~1 of \citealt{boy16}), and suggests that additional dip complexes may be seen in the future.  While the similarity of {\it Skara Brae} to the \kepler\ D1540 event (Figure~\ref{fig:popjupitercorn}) suggests a potential periodicity for this particular event, the broader differences between the \elsie\ family of dips and the Q16 complex suggest a significant stochastic component to the dip mechanism, if it is periodic.  The detection of dips from multiple independent observatories as illustrated in Figure~\ref{fig:phot_montage} also demonstrates they are real (i.e., astrophysical in origin), firmly ruling out an instrumental/data-reduction artifact as the source of the \kepler\ dips.

While they are far more precise, the \emph{Kepler} observations were in a single wide bandpass, so the color dependence of the dips could be inferred only with multiwavelength observations such as those presented here. These colors provide important new constraints; in a scenario where the dips are caused by material passing between us and the star, the colors measured in Section~\ref{sec:analysis} are consistent with extinction by optically thin submicron-sized dust (Figure~\ref{fig:opac}; see also Figure~3 of \citealt{wya17}). This conclusion is in contrast to longer-term observations that suggest that the wavelength dependence of the secular dimming is less pronounced, and therefore that intervening dust should be larger \citep{men17}. Thus, the current evidence suggests that the short-term and long-term dimming are caused by dust of different sizes. 

This difference could be a natural consequence of models that invoke exocomets or dust- enshrouded planetesimals; the dust concentrations that cause the short dips were created recently, and are richer in small, but short-lived, dust that is quickly ejected by radiation forces. Larger dust that is created survives and remains on a circumstellar orbit spreading from its point of origin in a manner similar to comet dust tails, causing the secular dimming.  In such a scenario, we also would not expect to see the same dust causing the short-duration dips in Figure~\ref{fig:phot_all} to return one orbit later, although the source of the dust may return one orbit later, creating fresh dust. 

If we further assume that the material is in a highly elliptical, circumstellar orbit, \citet{wya17} provide predictions to the infrared component of the dip based on several possible orbital configurations. In Section~\ref{sec:neowise}, our results show that the NEOWISE W1 and W2 measurements taken coincident with \elsie\ would rule out orbits with pericenter distances $q < 0.03$~au and $q > 0.3$~au.  If \elsie\ were a deeper dip, the constraints from the mid-infrared observations would have been stronger. However, in this case, the \citet{wya17} models show that to better constrain the location of material during a 2\% dip, observations taken at 10--30\,$\mu$m with a future space telescope would be necessary.

\elsie's dip depth wavelength dependence constrains and challenges many proffered models for the dips. For instance, intervening opaque material (e.g., a planet or megastructure) should produce wavelength-dependent dips only to the degree that there is nonuniformity across the stellar disk, such as from limb darkening or gravity darkening. Using a \citet{man02} model assuming a central transit and \citet{cla11} limb-darkening coefficients appropriate for \thisstar\ (and assuming negligible gravity darkening, appropriate given the star's $v\sin{i}$ value), we find a difference of 10\% between transit depths in notional filters around 4400\,\AA\ (approximately $B$) and 7600\,\AA\ (approximately $i^\prime$), implying a $B/i^\prime$ depth ratio of $\sim 1.1$, in strong disagreement with our observed values of $\sim 2$ (Section~\ref{sec:depths}; Figure~\ref{fig:Bayes}).  

To explore whether the observed colors are consistent with the appearance of cool spots, we have used both blackbody and \citet{Kurucz} model spectra to compute the brightness of a star in the $B$ and $i^\prime$ bands as functions of temperature $T$, where $T_0$ is the nominal effective temperature of \thisstar, 6720\,K.\footnote{We use the $B$ filter curve from \citet{Bes90} and the $i^\prime$ filter curve from \url{http://www.aip.de/en/research/facilities/stella/instruments/data/sloanugriz-filter-curves}.} We calculate the depth of a dip in band $X$ due to the temporary appearance of a spot of temperature $T$ occupying a fraction $A$ of the stellar surface as 
\begin{equation}
    \delta_X =\frac{X(T_0) - ((1-A)X(T_0) +A X(T))}{X(T_0)} = A\left(1-\frac{X(T)}{X(T_0)}\right)
\end{equation}
where (as with the data presented here) we normalize the brightness of the star against the pre-dip brightness. The ratio of the dip depths in the $B$ and $i^\prime$ bands in these models is then 
\begin{equation}
\frac{\delta_B}{\delta_{i^\prime}} = \frac{1-B(T)/B(T_0)}{1-i^\prime(T)/i^\prime(T_0)},
\end{equation}

\noindent independent of spot size (this model thus includes the case of the entire photosphere changing temperature at constant radius, which is more consistent with the lack of rotational modulation of any surface features). We find that this function yields dip depth ratios of 1.86 in the case of Kurucz model atmospheres and 1.65 with a blackbody approximation. The former value is formally 1.3$\sigma$ from our measured value of $B/i^\prime = 1.94\pm 0.06$, however only 7\% of our posterior samples had $B/i^\prime < 1.86$. Our {\it Elsie} multiband photometry thus may be consistent with models in which transient cool surface regions explain the dips.\footnote{Our analysis are also in tension with that of \citet{Foukal17} on this same data set, who finds the $B/i^\prime$ dip depth ratio for {\it Elsie} to be closer to unity and more consistent with the blackbody expectation.}  Note that this simple analysis does not address models invoking the {\it dis}appearance of regions much hotter than $T_0$, or those that invoke a changing stellar radius. 

Invoking dust still challenges our creativity in developing a unified theory to explain all the observations; however, the models of \citet{wya17} give hope to a swarm of yet unspecified objects in an eccentric orbit (in this case, exocomets, with an alternative being dust-enshrouded planetesimals as proposed by \citealt{nes17}) causing the brightness fluctuations.  Continued monitoring to detect events in the future will help narrow down any periodicity within the dip occurrence, which would strengthen the argument that the source of the obscuring material was in orbit around the star, as opposed to density fluctuations in the ISM, etc.  

In fact, if the two deepest events in the \kepler\ data were caused by a recent planet-planet collision, \citet{boy16} predicted 2017 May to be the next event -- and this was precisely when \elsie\ appeared. However, this prediction was based solely on the two large events taking place roughly two years apart (in 2011 and again in 2013), and came with several caveats. The obvious issue discussed was the low probability of witnessing such an event, given the lifetime of the \kepler\ mission, geometry, system age, and lack of any IR excess based on the timing of the WISE observations (circa 2010).  Another issue with this prediction, however, is that it would not explain the \kepler\ dips that happened between the large events.  Moreover, even though the \elsie\ family appeared timely in this prediction, the other predicted 2015 April event went undetected because no monitoring was ongoing at the time\footnote{There was no concerted monitoring effort of the star from 2013 May (when \kepler\ discontinued observations) to 2015 October.}.  This gap in time coverage hinders any truly periodic interpretation based on the occurrence {\it and} nonoccurrence of the dips\footnote{Monitoring from 2016 March up until \elsie\ by LCOGT showed no events at the $>1$\% level.  Monitoring from 2015 October up until \elsie\ by the American Association of Variable Star Observers (AAVSO) showed no events at the $>2$\% level. These observations will be presented in a forthcoming paper.}. Furthermore, as previously mentioned, each of the dips lack resemblance to one another, and while nonstatic shape and orientation would be expected if the material is continuously being pulverized as it orbits the star, and/or if the newly formed small-dust particle concentrations get ejected quickly due to radiation pressure as discussed above, this quality makes matching the \elsie\ family and the \kepler\ Q16 complex a challenging task.  To that end, \citet{sac17} point out that the \elsie\ family dips (Figure~\ref{fig:phot_all}) occur in an interval perhaps reminiscent of the 48.4\,day putative period discussed by \citet{boy16} for the \kepler\ dip times, leading to a prediction of a 1574\,day period (or a semimajor axis of 2.6~au), roughly double the period proposed by \citet{boy16}.  Observations in June 2019 (750 days from 2017 May) will be critical to determine whether the \kepler\ D800 event \citep{boy16} repeats.

\section{Conclusions}

Given the \kepler\ data alone, it proved difficult to study this star because ground-based follow-up observations were not taken contemporaneously with the dipping events. 
In this paper, we show that we are able to successfully trigger a worldwide request for observations using a variety of telescopes and instruments, with different techniques, sensitivities, resolutions, and wavelengths. 
The main results within this paper include:
\begin{itemize}
    \item The photometric monitoring of \thisstar\ is the first successful effort via crowd-funding to study an astronomical object.
    \item We present the {\it Elsie} family of dips, a series of 1--2.5\% dips that began in 2017 May and carried on through the end of December, at the time of this paper's writing.  The {\it Elsie} family consists of four main dipping events, {\it ``Elsie,'' ``Celeste,'' ``Skara Brae,''}, and {\it ``Angkor''} all which last for several days to weeks. 
    \item Our observations mark the first real-time detection of a dip in brightness for \thisstar.  Triggered spectroscopic and polarmetric observations taken during the dips reveal no large, obvious changes compared to out of dip observations. 
    \item Multiband photometry taken during {\it Elsie} show its amplitude is chromatic, with depth ratios that are consistent with occultation by optically thin dust with size scales $\ll 1 \mu$m, and perhaps with variations intrinsic to the star.
\end{itemize}

\thisstar\ has captured the imagination of both scientists and the public.  To that end, our team strives to make the steps taken to learn more about the star as transparent as possible.  Additional constraints on the system will come from the triggered observations taken during the \elsie\ family of dips and beyond, which will in turn allow for more detailed modeling. Opportunities include observational projects from numerous facilities, impressively demonstrating the multidimensional approach of the community to study \thisstar, as mentioned within the above sections.  
The observed ``colors'' of the dips (i.e.\ the ratios of the dip depths in different bands) appear inconsistent with occultation by primarily optically thick material (which would be expected to produce nearly achromatic dips) and appear to be in some tension with intrinsic cooling of the star at constant radius.

We emphasize the importance that continued monitoring will bring to our understanding of the physical processes responsible for the light curve features.  In general, precise, long-term photometric monitoring to detect future dips is a level-zero requirement.  These data also provide the means of informing planned triggered observations such as high-resolution spectroscopy to study the events in more detail. Furthermore, extended photometric monitoring will enable us to characterize the star's long-term variability \citep{sch16,mon16,men17,sim17}, which is thought to be linked to the dips in some way.  All-in-all, the apparent low duty cycle of the dips, unclear predictions on when they will recur, and fairly unconstrained multiyear timescales of the long-term variability will require a committed, intensive monitoring program spanning the next decade and beyond.

\acknowledgements{
The LCOGT observations used in this project were made possible by the collective effort of 1,762 supporters as part of the Kickstarter campaign ``The Most Mysterious Star in the Galaxy''\footnote{\url{https://www.kickstarter.com/projects/608159144/the-most-mysterious-star-in-the-galaxy/}}. The authors gratefully acknowledge and humbly extend a special thanks for substantial support from Las Cumbres Observatory, Glenn Klakring, Fred Boyajian \& Bobbie Staley, Alex Mazingue, The Bible Family, Claudio Bottaccini, Joachim De Lombaert, Amity \& Brigid Williams, Kevin Fischer, William Hopkins, Milton Bosch, Zipeng Wang, TJ, DR, and CC. The authors thank Peter Foukal for helpful conversations.
A. Tanner wishes to acknowledge her mother, Celeste, for all her love and efforts
to help her become an astronomer. A. Tanner also wishes to thank the Boyajian's Star
Kickstarter team for electing to name dip {\it ``Celeste''} after her mom who passed
away in June 2017. Mom would have been pleased and would have ``got it.'' 

H.D. acknowledges support from grant ESP2015-65712-C5-4-R of the Spanish Secretary of State for R\&D\&i (MINECO). This research made use of data acquired with the Gran Telescopio Canarias (GTC), installed at the Spanish Observatorio del Roque de los Muchachos of the Instituto de Astrof\'\i sica de Canarias, in the island of La Palma.
JB acknowledges the support from the VEGA 2/0031/18 and APVV 15-0458 grants.
The Center for Exoplanets and Habitable Worlds is supported by the Pennsylvania State
University, the Eberly College of Science, and the Pennsylvania Space Grant Consortium. 
E.H.L.B.'s research was supported by an appointment to the NASA Postdoctoral Program
with the Nexus for Exoplanet System Science, administered by Universities Space Research
Association under contract with NASA.
A.V.F.'s group is grateful for financial assistance from the TABASGO Foundation,
the Christopher R. Redlich Fund, Gary and Cynthia Bengier (T.d.J. is a Bengier
Postdoctoral fellow), and the Miller Institute for Basic Research in Science (U.C.
Berkeley).
This project was supported by the National Research, Development and Innovation Fund of
Hungary, financed under the K\_16 funding scheme (project No. NKFIH K-115709).
The research leading to these results has received funding from the ARC grant for
Concerted Research Actions, financed by the Wallonia-Brussels Federation. M. Gillon is
Research Associate at the Belgian Fonds de la Recherche Scientifique (F.R.S.-FNRS).

A major upgrade of the Kast spectrograph on the Shane 3~m telescope
at Lick Observatory was made possible through generous gifts from
William and Marina Kast as well as the Heising-Simons Foundation.
Research at Lick Observatory is partially supported by a generous
gift from Google.
Some of the data presented herein were obtained at the W. M. Keck
Observatory, which is operated as a scientific partnership among the
California Institute of Technology, the University of California, and the
National Aeronautics and Space Administration
(NASA); the observatory was made possible by the generous financial
support of the W. M. Keck Foundation.
This publication makes use of data products from the Wide-field Infrared Survey
Explorer, which is a joint project of the University of California, Los Angeles, and the
Jet Propulsion Laboratory/California Institute of Technology, funded by NASA.  This publication also makes use of data products
from NEOWISE, which is a project of the Jet Propulsion Laboratory/California Institute
of Technology, funded by the Planetary Science Division of NASA.

}

\software{emcee \citep{emcee13}}

\clearpage
\bibliographystyle{apj}            

\begin{thebibliography}{0}
\expandafter\ifx\csname natexlab\endcsname\relax\def\natexlab#1{#1}\fi

\end{thebibliography}


\begin{thebibliography}{69}
\expandafter\ifx\csname natexlab\endcsname\relax\def\natexlab#1{#1}\fi

\bibitem[{{Ballesteros} {et~al.}(2017){Ballesteros}, {Arnalte-Mur},
  {Fernandez-Soto}, \& {Martinez}}]{bal17}
{Ballesteros}, F.~J., {Arnalte-Mur}, P., {Fernandez-Soto}, A., \& {Martinez},
  V.~J. 2017, ArXiv e-prints

\bibitem[{{Barbary}(2016)}]{bar16}
{Barbary}, K. 2016, {SEP: Source Extractor as a library}

\bibitem[{{Bertin} \& {Arnouts}(1996)}]{ber96}
{Bertin}, E. \& {Arnouts}, S. 1996, \aaps, 117, 393

\bibitem[{{Bessell}(1990)}]{Bes90}
{Bessell}, M.~S. 1990, \pasp, 102, 1181

\bibitem[{{Bodman} \& {Quillen}(2016)}]{bod16}
{Bodman}, E.~H.~L. \& {Quillen}, A. 2016, \apjl, 819, L34

\bibitem[{{Borucki} {et~al.}(2010){Borucki}, {Koch}, {Basri}, {Batalha},
  {Brown}, {Caldwell}, {Caldwell}, {Christensen-Dalsgaard}, {Cochran},
  {DeVore}, {Dunham}, {Dupree}, {Gautier}, {Geary}, {Gilliland}, {Gould},
  {Howell}, {Jenkins}, {Kondo}, {Latham}, {Marcy}, {Meibom}, {Kjeldsen},
  {Lissauer}, {Monet}, {Morrison}, {Sasselov}, {Tarter}, {Boss}, {Brownlee},
  {Owen}, {Buzasi}, {Charbonneau}, {Doyle}, {Fortney}, {Ford}, {Holman},
  {Seager}, {Steffen}, {Welsh}, {Rowe}, {Anderson}, {Buchhave}, {Ciardi},
  {Walkowicz}, {Sherry}, {Horch}, {Isaacson}, {Everett}, {Fischer}, {Torres},
  {Johnson}, {Endl}, {MacQueen}, {Bryson}, {Dotson}, {Haas}, {Kolodziejczak},
  {Van Cleve}, {Chandrasekaran}, {Twicken}, {Quintana}, {Clarke}, {Allen},
  {Li}, {Wu}, {Tenenbaum}, {Verner}, {Bruhweiler}, {Barnes}, \& {Prsa}}]{bor10}
{Borucki}, W.~J., {Koch}, D., {Basri}, G., {Batalha}, N., {Brown}, T.,
  {Caldwell}, D., {Caldwell}, J., {Christensen-Dalsgaard}, J., {Cochran},
  W.~D., {DeVore}, E., {Dunham}, E.~W., {Dupree}, A.~K., {Gautier}, T.~N.,
  {Geary}, J.~C., {Gilliland}, R., {Gould}, A., {Howell}, S.~B., {Jenkins},
  J.~M., {Kondo}, Y., {Latham}, D.~W., {Marcy}, G.~W., {Meibom}, S.,
  {Kjeldsen}, H., {Lissauer}, J.~J., {Monet}, D.~G., {Morrison}, D.,
  {Sasselov}, D., {Tarter}, J., {Boss}, A., {Brownlee}, D., {Owen}, T.,
  {Buzasi}, D., {Charbonneau}, D., {Doyle}, L., {Fortney}, J., {Ford}, E.~B.,
  {Holman}, M.~J., {Seager}, S., {Steffen}, J.~H., {Welsh}, W.~F., {Rowe}, J.,
  {Anderson}, H., {Buchhave}, L., {Ciardi}, D., {Walkowicz}, L., {Sherry}, W.,
  {Horch}, E., {Isaacson}, H., {Everett}, M.~E., {Fischer}, D., {Torres}, G.,
  {Johnson}, J.~A., {Endl}, M., {MacQueen}, P., {Bryson}, S.~T., {Dotson}, J.,
  {Haas}, M., {Kolodziejczak}, J., {Van Cleve}, J., {Chandrasekaran}, H.,
  {Twicken}, J.~D., {Quintana}, E.~V., {Clarke}, B.~D., {Allen}, C., {Li}, J.,
  {Wu}, H., {Tenenbaum}, P., {Verner}, E., {Bruhweiler}, F., {Barnes}, J., \&
  {Prsa}, A. 2010, Science, 327, 977

\bibitem[{{Boyajian} {et~al.}(2016){Boyajian}, {LaCourse}, {Rappaport},
  {Fabrycky}, {Fischer}, {Gandolfi}, {Kennedy}, {Korhonen}, {Liu}, {Moor},
  {Olah}, {Vida}, {Wyatt}, {Best}, {Brewer}, {Ciesla}, {Cs{\'a}k}, {Deeg},
  {Dupuy}, {Handler}, {Heng}, {Howell}, {Ishikawa}, {Kov{\'a}cs}, {Kozakis},
  {Kriskovics}, {Lehtinen}, {Lintott}, {Lynn}, {Nespral}, {Nikbakhsh},
  {Schawinski}, {Schmitt}, {Smith}, {Szabo}, {Szabo}, {Viuho}, {Wang},
  {Weiksnar}, {Bosch}, {Connors}, {Goodman}, {Green}, {Hoekstra}, {Jebson},
  {Jek}, {Omohundro}, {Schwengeler}, \& {Szewczyk}}]{boy16}
{Boyajian}, T.~S., {LaCourse}, D.~M., {Rappaport}, S.~A., {Fabrycky}, D.,
  {Fischer}, D.~A., {Gandolfi}, D., {Kennedy}, G.~M., {Korhonen}, H., {Liu},
  M.~C., {Moor}, A., {Olah}, K., {Vida}, K., {Wyatt}, M.~C., {Best}, W.~M.~J.,
  {Brewer}, J., {Ciesla}, F., {Cs{\'a}k}, B., {Deeg}, H.~J., {Dupuy}, T.~J.,
  {Handler}, G., {Heng}, K., {Howell}, S.~B., {Ishikawa}, S.~T., {Kov{\'a}cs},
  J., {Kozakis}, T., {Kriskovics}, L., {Lehtinen}, J., {Lintott}, C., {Lynn},
  S., {Nespral}, D., {Nikbakhsh}, S., {Schawinski}, K., {Schmitt}, J.~R.,
  {Smith}, A.~M., {Szabo}, G., {Szabo}, R., {Viuho}, J., {Wang}, J.,
  {Weiksnar}, A., {Bosch}, M., {Connors}, J.~L., {Goodman}, S., {Green}, G.,
  {Hoekstra}, A.~J., {Jebson}, T., {Jek}, K.~J., {Omohundro}, M.~R.,
  {Schwengeler}, H.~M., \& {Szewczyk}, A. 2016, \mnras, 457, 3988

\bibitem[{{Brown} {et~al.}(2013){Brown}, {Baliber}, {Bianco}, {Bowman},
  {Burleson}, {Conway}, {Crellin}, {Depagne}, {De Vera}, {Dilday}, {Dragomir},
  {Dubberley}, {Eastman}, {Elphick}, {Falarski}, {Foale}, {Ford}, {Fulton},
  {Garza}, {Gomez}, {Graham}, {Greene}, {Haldeman}, {Hawkins}, {Haworth},
  {Haynes}, {Hidas}, {Hjelstrom}, {Howell}, {Hygelund}, {Lister}, {Lobdill},
  {Martinez}, {Mullins}, {Norbury}, {Parrent}, {Paulson}, {Petry}, {Pickles},
  {Posner}, {Rosing}, {Ross}, {Sand}, {Saunders}, {Shobbrook}, {Shporer},
  {Street}, {Thomas}, {Tsapras}, {Tufts}, {Valenti}, {Vander Horst}, {Walker},
  {White}, \& {Willis}}]{bro13}
{Brown}, T.~M., {Baliber}, N., {Bianco}, F.~B., {Bowman}, M., {Burleson}, B.,
  {Conway}, P., {Crellin}, M., {Depagne}, {\'E}., {De Vera}, J., {Dilday}, B.,
  {Dragomir}, D., {Dubberley}, M., {Eastman}, J.~D., {Elphick}, M., {Falarski},
  M., {Foale}, S., {Ford}, M., {Fulton}, B.~J., {Garza}, J., {Gomez}, E.~L.,
  {Graham}, M., {Greene}, R., {Haldeman}, B., {Hawkins}, E., {Haworth}, B.,
  {Haynes}, R., {Hidas}, M., {Hjelstrom}, A.~E., {Howell}, D.~A., {Hygelund},
  J., {Lister}, T.~A., {Lobdill}, R., {Martinez}, J., {Mullins}, D.~S.,
  {Norbury}, M., {Parrent}, J., {Paulson}, R., {Petry}, D.~L., {Pickles}, A.,
  {Posner}, V., {Rosing}, W.~E., {Ross}, R., {Sand}, D.~J., {Saunders}, E.~S.,
  {Shobbrook}, J., {Shporer}, A., {Street}, R.~A., {Thomas}, D., {Tsapras}, Y.,
  {Tufts}, J.~R., {Valenti}, S., {Vander Horst}, K., {Walker}, Z., {White}, G.,
  \& {Willis}, M. 2013, \pasp, 125, 1031

\bibitem[{{Budaj} {et~al.}(2015){Budaj}, {Kocifaj}, {Salmeron}, \&
  {Hubeny}}]{budaj15}
{Budaj}, J., {Kocifaj}, M., {Salmeron}, R., \& {Hubeny}, I. 2015, \mnras, 454,
  2

\bibitem[{{Castelli} \& {Kurucz}(2004)}]{Kurucz}
{Castelli}, F. \& {Kurucz}, R.~L. 2004, arxiv:astro-ph/0405087

\bibitem[{{Claret} \& {Bloemen}(2011)}]{cla11}
{Claret}, A. \& {Bloemen}, S. 2011, \aap, 529, A75

\bibitem[{{Collins} {et~al.}(2017){Collins}, {Kielkopf}, {Stassun}, \&
  {Hessman}}]{col17}
{Collins}, K.~A., {Kielkopf}, J.~F., {Stassun}, K.~G., \& {Hessman}, F.~V.
  2017, \aj, 153, 77

\bibitem[{{Colome} \& {Ribas}(2006)}]{col06}
{Colome}, J. \& {Ribas}, I. 2006, IAU Special Session, 6

\bibitem[{{Craig} {et~al.}(2015){Craig}, {Crawford}, {Deil}, {Gomez},
  {G{\"u}nther}, {Heidt}, {Horton}, {Karr}, {Nelson}, {Ninan}, {Pattnaik},
  {Rol}, {Schoenell}, {Seifert}, {Singh}, {Sipocz}, {Stotts}, {Streicher},
  {Tollerud}, {Walker}, \& {ccdproc Contributors}}]{cra15}
{Craig}, M.~W., {Crawford}, S.~M., {Deil}, C., {Gomez}, C., {G{\"u}nther},
  H.~M., {Heidt}, N., {Horton}, A., {Karr}, J., {Nelson}, S., {Ninan}, J.~P.,
  {Pattnaik}, P., {Rol}, E., {Schoenell}, W., {Seifert}, M., {Singh}, S.,
  {Sipocz}, B., {Stotts}, C., {Streicher}, O., {Tollerud}, E., {Walker}, N., \&
  {ccdproc Contributors}. 2015, {ccdproc: CCD data reduction software},
  Astrophysics Source Code Library

\bibitem[{{Curtis}(2017)}]{Curtis2017}
{Curtis}, J.~L. 2017, \aj, 153, 275

\bibitem[{{Cutri} {et~al.}(2015){Cutri}, {Mainzer}, {Conrow}, {Masci}, {Bauer},
  {Dailey}, {Kirkpatrick}, {Fajardo-Acosta}, {Gelino}, {Grillmair}, {Wheelock},
  {Yan}, {Harbut}, {Beck}, {Wittman}, {Wright}, {Masiero}, {Grav}, {Sonnett},
  {Nugent}, {Kramer}, {Stevenson}, {Eisenhardt}, {Fabinsky}, {Tholen}, {Papin},
  {Fowler}, \& {McCallon}}]{cut15}
{Cutri}, R.~M., {Mainzer}, A., {Conrow}, T., {Masci}, F., {Bauer}, J.,
  {Dailey}, J., {Kirkpatrick}, J.~D., {Fajardo-Acosta}, S., {Gelino}, C.,
  {Grillmair}, C., {Wheelock}, S.~L., {Yan}, L., {Harbut}, M., {Beck}, R.,
  {Wittman}, M., {Wright}, E.~L., {Masiero}, J., {Grav}, T., {Sonnett}, S.,
  {Nugent}, C., {Kramer}, E., {Stevenson}, R., {Eisenhardt}, P.~R.~M.,
  {Fabinsky}, B., {Tholen}, D., {Papin}, M., {Fowler}, J., \& {McCallon}, H.
  2015, {Explanatory Supplement to the NEOWISE Data Release Products}, Tech.
  rep.

\bibitem[{{Deirmendjian}(1964)}]{dei64}
{Deirmendjian}, D. 1964, \ao, 3, 187

\bibitem[{{Dorschner} {et~al.}(1995){Dorschner}, {Begemann}, {Henning},
  {Jaeger}, \& {Mutschke}}]{Dorschner95}
{Dorschner}, J., {Begemann}, B., {Henning}, T., {Jaeger}, C., \& {Mutschke}, H.
  1995, \aap, 300, 503

\bibitem[{{Faber} {et~al.}(2003){Faber}, {Phillips}, {Kibrick}, {Alcott},
  {Allen}, {Burrous}, {Cantrall}, {Clarke}, {Coil}, {Cowley}, {Davis}, {Deich},
  {Dietsch}, {Gilmore}, {Harper}, {Hilyard}, {Lewis}, {McVeigh}, {Newman},
  {Osborne}, {Schiavon}, {Stover}, {Tucker}, {Wallace}, {Wei}, {Wirth}, \&
  {Wright}}]{fab03}
{Faber}, S.~M., {Phillips}, A.~C., {Kibrick}, R.~I., {Alcott}, B., {Allen},
  S.~L., {Burrous}, J., {Cantrall}, T., {Clarke}, D., {Coil}, A.~L., {Cowley},
  D.~J., {Davis}, M., {Deich}, W.~T.~S., {Dietsch}, K., {Gilmore}, D.~K.,
  {Harper}, C.~A., {Hilyard}, D.~F., {Lewis}, J.~P., {McVeigh}, M., {Newman},
  J., {Osborne}, J., {Schiavon}, R., {Stover}, R.~J., {Tucker}, D., {Wallace},
  V., {Wei}, M., {Wirth}, G., \& {Wright}, C.~A. 2003, in \procspie, Vol. 4841,
  Instrument Design and Performance for Optical/Infrared Ground-based
  Telescopes, ed. M.~{Iye} \& A.~F.~M. {Moorwood}, 1657--1669

\bibitem[{{Filippenko}(1982)}]{fil82}
{Filippenko}, A.~V. 1982, \pasp, 94, 715

\bibitem[{{Foreman-Mackey} {et~al.}(2013){Foreman-Mackey}, {Hogg}, {Lang}, \&
  {Goodman}}]{emcee13}
{Foreman-Mackey}, D., {Hogg}, D.~W., {Lang}, D., \& {Goodman}, J. 2013, \pasp,
  125, 306

\bibitem[{{Foukal}(2017)}]{fou17}
{Foukal}, P. 2017, \apjl, 842, L3

\bibitem[{Foukal(2017)}]{Foukal17}
Foukal, P. 2017, Research Notes of the AAS, 1, 52

\bibitem[{{Gillon} {et~al.}(2011){Gillon}, {Jehin}, {Magain}, {Chantry},
  {Hutsem{\'e}kers}, {Manfroid}, {Queloz}, \& {Udry}}]{Gil11}
{Gillon}, M., {Jehin}, E., {Magain}, P., {Chantry}, V., {Hutsem{\'e}kers}, D.,
  {Manfroid}, J., {Queloz}, D., \& {Udry}, S. 2011, in European Physical
  Journal Web of Conferences, Vol.~11, European Physical Journal Web of
  Conferences, 06002

\bibitem[{{Harp} {et~al.}(2016){Harp}, {Richards}, {Shostak}, {Tarter},
  {Vakoch}, \& {Munson}}]{har16}
{Harp}, G.~R., {Richards}, J., {Shostak}, S., {Tarter}, J.~C., {Vakoch}, D.~A.,
  \& {Munson}, C. 2016, \apj, 825, 155

\bibitem[{{Heiles}(2000)}]{hei00}
{Heiles}, C. 2000, \aj, 119, 923

\bibitem[{{Hippke} {et~al.}(2016){Hippke}, {Angerhausen}, {Lund}, {Pepper}, \&
  {Stassun}}]{hip16}
{Hippke}, M., {Angerhausen}, D., {Lund}, M.~B., {Pepper}, J., \& {Stassun},
  K.~G. 2016, \apj, 825, 73

\bibitem[{{J{\"a}ger} {et~al.}(2003){J{\"a}ger}, {Dorschner}, {Mutschke},
  {Posch}, \& {Henning}}]{jager03}
{J{\"a}ger}, C., {Dorschner}, J., {Mutschke}, H., {Posch}, T., \& {Henning}, T.
  2003, \aap, 408, 193

\bibitem[{{Jager} {et~al.}(1998){Jager}, {Mutschke}, \& {Henning}}]{jager98}
{Jager}, C., {Mutschke}, H., \& {Henning}, T. 1998, \aap, 332, 291

\bibitem[{{Jehin} {et~al.}(2011){Jehin}, {Gillon}, {Queloz}, {Magain},
  {Manfroid}, {Chantry}, {Lendl}, {Hutsem{\'e}kers}, \& {Udry}}]{Jeh11}
{Jehin}, E., {Gillon}, M., {Queloz}, D., {Magain}, P., {Manfroid}, J.,
  {Chantry}, V., {Lendl}, M., {Hutsem{\'e}kers}, D., \& {Udry}, S. 2011, The
  Messenger, 145, 2

\bibitem[{{Johnson} \& {Christy}(1974)}]{johnson74}
{Johnson}, P.~B. \& {Christy}, R.~W. 1974, \prb, 9, 5056

\bibitem[{{Katz}(2017)}]{kat17}
{Katz}, J.~I. 2017, \mnras, 471, 3680

\bibitem[{{Kiefer} {et~al.}(2017){Kiefer}, {Lecavelier des Etangs},
  {Vidal-Madjar}, {H{\'e}brard}, {Bourrier}, \& {Wilson}}]{kie17}
{Kiefer}, F., {Lecavelier des Etangs}, A., {Vidal-Madjar}, A., {H{\'e}brard},
  G., {Bourrier}, V., \& {Wilson}, P.-A. 2017, ArXiv e-prints

\bibitem[{{Koike} {et~al.}(1995){Koike}, {Kaito}, {Yamamoto}, {Shibai},
  {Kimura}, \& {Suto}}]{Koike95}
{Koike}, C., {Kaito}, C., {Yamamoto}, T., {Shibai}, H., {Kimura}, S., \&
  {Suto}, H. 1995, \icarus, 114, 203

\bibitem[{{Lisse} {et~al.}(2015){Lisse}, {Sitko}, \& {Marengo}}]{lis15}
{Lisse}, C.~M., {Sitko}, M.~L., \& {Marengo}, M. 2015, \apjl, 815, L27

\bibitem[{{Makarov} \& {Goldin}(2016)}]{mak16}
{Makarov}, V.~V. \& {Goldin}, A. 2016, \apj, 833, 78

\bibitem[{{Mandel} \& {Agol}(2002)}]{man02}
{Mandel}, K. \& {Agol}, E. 2002, \apjl, 580, L171

\bibitem[{{Marengo} {et~al.}(2015){Marengo}, {Hulsebus}, \& {Willis}}]{mar15}
{Marengo}, M., {Hulsebus}, A., \& {Willis}, S. 2015, \apjl, 814, L15

\bibitem[{{Mauerhan} {et~al.}(2014){Mauerhan}, {Williams}, {Smith}, {Smith},
  {Filippenko}, {Hoffman}, {Milne}, {Leonard}, {Clubb}, {Fox}, \&
  {Kelly}}]{mau14}
{Mauerhan}, J., {Williams}, G.~G., {Smith}, N., {Smith}, P.~S., {Filippenko},
  A.~V., {Hoffman}, J.~L., {Milne}, P., {Leonard}, D.~C., {Clubb}, K.~I.,
  {Fox}, O.~D., \& {Kelly}, P.~L. 2014, \mnras, 442, 1166

\bibitem[{{McCormac} {et~al.}(2014){McCormac}, {Skillen}, {Pollacco}, {Faedi},
  {Ramsay}, {Dhillon}, {Todd}, \& {Gonzalez}}]{mcc14}
{McCormac}, J., {Skillen}, I., {Pollacco}, D., {Faedi}, F., {Ramsay}, G.,
  {Dhillon}, V.~S., {Todd}, I., \& {Gonzalez}, A. 2014, \mnras, 438, 3383

\bibitem[{{Meng} {et~al.}(2017){Meng}, {Rieke}, {Dubois}, {Kennedy}, {Marengo},
  {Siegel}, {Su}, {Trueba}, {Wyatt}, {Boyajian}, {Lisse}, {Logie}, {Rau}, \&
  {Vanaverbeke}}]{men17}
{Meng}, H.~Y.~A., {Rieke}, G., {Dubois}, F., {Kennedy}, G., {Marengo}, M.,
  {Siegel}, M., {Su}, K., {Trueba}, N., {Wyatt}, M., {Boyajian}, T., {Lisse},
  C.~M., {Logie}, L., {Rau}, S., \& {Vanaverbeke}, S. 2017, \apj, 847, 131

\bibitem[{{Metzger} {et~al.}(2017){Metzger}, {Shen}, \& {Stone}}]{met17}
{Metzger}, B.~D., {Shen}, K.~J., \& {Stone}, N. 2017, \mnras, 468, 4399

\bibitem[{{Miller} \& {Stone}(2013)}]{miller-stone93}
{Miller}, J.~S. \& {Stone}, R. P.~S. 2013, Lick Obs. Tech. Rep., 66

\bibitem[{{Molnar} {et~al.}(2017){Molnar}, {Van Noord}, {Kinemuchi},
  {Smolinski}, {Alexander}, {Cook}, {Jang}, {Kobulnicky}, {Spedden}, \&
  {Steenwyk}}]{mol17}
{Molnar}, L.~A., {Van Noord}, D.~M., {Kinemuchi}, K., {Smolinski}, J.~P.,
  {Alexander}, C.~E., {Cook}, E.~M., {Jang}, B., {Kobulnicky}, H.~A.,
  {Spedden}, C.~J., \& {Steenwyk}, S.~D. 2017, \apj, 840, 1

\bibitem[{{Montet} \& {Simon}(2016)}]{mon16}
{Montet}, B.~T. \& {Simon}, J.~D. 2016, \apjl, 830, L39

\bibitem[{{Neslu{\v s}an} \& {Budaj}(2017)}]{nes17}
{Neslu{\v s}an}, L. \& {Budaj}, J. 2017, \aap, 600, A86

\bibitem[{{Oke} {et~al.}(1995){Oke}, {Cohen}, {Carr}, {Cromer}, {Dingizian},
  {Harris}, {Labrecque}, {Lucinio}, {Schaal}, {Epps}, \& {Miller}}]{oke95}
{Oke}, J.~B., {Cohen}, J.~G., {Carr}, M., {Cromer}, J., {Dingizian}, A.,
  {Harris}, F.~H., {Labrecque}, S., {Lucinio}, R., {Schaal}, W., {Epps}, H., \&
  {Miller}, J. 1995, \pasp, 107, 375

\bibitem[{{Pecaut} \& {Mamajek}(2013)}]{pec13}
{Pecaut}, M.~J. \& {Mamajek}, E.~E. 2013, \apjs, 208, 9

\bibitem[{{Pecaut} {et~al.}(2012){Pecaut}, {Mamajek}, \& {Bubar}}]{pec12}
{Pecaut}, M.~J., {Mamajek}, E.~E., \& {Bubar}, E.~J. 2012, \apj, 746, 154

\bibitem[{{Rockosi} {et~al.}(2010){Rockosi}, {Stover}, {Kibrick}, {Lockwood},
  {Peck}, {Cowley}, {Bolte}, {Adkins}, {Alcott}, {Allen}, {Brown}, {Cabak},
  {Deich}, {Hilyard}, {Kassis}, {Lanclos}, {Lewis}, {Pfister}, {Phillips},
  {Robinson}, {Saylor}, {Thompson}, {Ward}, {Wei}, \& {Wright}}]{roc10}
{Rockosi}, C., {Stover}, R., {Kibrick}, R., {Lockwood}, C., {Peck}, M.,
  {Cowley}, D., {Bolte}, M., {Adkins}, S., {Alcott}, B., {Allen}, S.~L.,
  {Brown}, B., {Cabak}, G., {Deich}, W., {Hilyard}, D., {Kassis}, M.,
  {Lanclos}, K., {Lewis}, J., {Pfister}, T., {Phillips}, A., {Robinson}, L.,
  {Saylor}, M., {Thompson}, M., {Ward}, J., {Wei}, M., \& {Wright}, C. 2010, in
  \procspie, Vol. 7735, Ground-based and Airborne Instrumentation for Astronomy
  III, 77350R

\bibitem[{{Sacco} {et~al.}(2017){Sacco}, {Ngo}, \& {Modolo}}]{sac17}
{Sacco}, G., {Ngo}, L., \& {Modolo}, J. 2017, ArXiv e-prints

\bibitem[{{Savage} \& {Mathis}(1979)}]{sav79}
{Savage}, B.~D. \& {Mathis}, J.~S. 1979, \araa, 17, 73

\bibitem[{{Schaefer}(2016)}]{sch16}
{Schaefer}, B.~E. 2016, \apjl, 822, L34

\bibitem[{{Schmidt} {et~al.}(1992){Schmidt}, {Elston}, \& {Lupie}}]{sch92}
{Schmidt}, G.~D., {Elston}, R., \& {Lupie}, O.~L. 1992, \aj, 104, 1563

\bibitem[{{Serkowski} {et~al.}(1975){Serkowski}, {Mathewson}, \&
  {Ford}}]{ser75}
{Serkowski}, K., {Mathewson}, D.~S., \& {Ford}, V.~L. 1975, \apj, 196, 261

\bibitem[{{Sheikh} {et~al.}(2016){Sheikh}, {Weaver}, \& {Dahmen}}]{she16}
{Sheikh}, M.~A., {Weaver}, R.~L., \& {Dahmen}, K.~A. 2016, Physical Review
  Letters, 117, 261101

\bibitem[{{Silverman} {et~al.}(2012){Silverman}, {Foley}, {Filippenko},
  {Ganeshalingam}, {Barth}, {Chornock}, {Griffith}, {Kong}, {Lee}, {Leonard},
  {Matheson}, {Miller}, {Steele}, {Barris}, {Bloom}, {Cobb}, {Coil},
  {Desroches}, {Gates}, {Ho}, {Jha}, {Kandrashoff}, {Li}, {Mandel}, {Modjaz},
  {Moore}, {Mostardi}, {Papenkova}, {Park}, {Perley}, {Poznanski}, {Reuter},
  {Scala}, {Serduke}, {Shields}, {Swift}, {Tonry}, {Van Dyk}, {Wang}, \&
  {Wong}}]{sil12}
{Silverman}, J.~M., {Foley}, R.~J., {Filippenko}, A.~V., {Ganeshalingam}, M.,
  {Barth}, A.~J., {Chornock}, R., {Griffith}, C.~V., {Kong}, J.~J., {Lee}, N.,
  {Leonard}, D.~C., {Matheson}, T., {Miller}, E.~G., {Steele}, T.~N., {Barris},
  B.~J., {Bloom}, J.~S., {Cobb}, B.~E., {Coil}, A.~L., {Desroches}, L.-B.,
  {Gates}, E.~L., {Ho}, L.~C., {Jha}, S.~W., {Kandrashoff}, M.~T., {Li}, W.,
  {Mandel}, K.~S., {Modjaz}, M., {Moore}, M.~R., {Mostardi}, R.~E.,
  {Papenkova}, M.~S., {Park}, S., {Perley}, D.~A., {Poznanski}, D., {Reuter},
  C.~A., {Scala}, J., {Serduke}, F.~J.~D., {Shields}, J.~C., {Swift}, B.~J.,
  {Tonry}, J.~L., {Van Dyk}, S.~D., {Wang}, X., \& {Wong}, D.~S. 2012, \mnras,
  425, 1789

\bibitem[{{Simon} {et~al.}(2017){Simon}, {Shappee}, {Pojmanski}, {Montet},
  {Kochanek}, {van Saders}, {Holoien}, \& {Henden}}]{sim17}
{Simon}, J.~D., {Shappee}, B.~J., {Pojmanski}, G., {Montet}, B.~T., {Kochanek},
  C.~S., {van Saders}, J., {Holoien}, T.~W.-S., \& {Henden}, A.~A. 2017, ArXiv
  e-prints

\bibitem[{{Sing} {et~al.}(2015){Sing}, {Wakeford}, {Showman}, {Nikolov},
  {Fortney}, {Burrows}, {Ballester}, {Deming}, {Aigrain}, {D{\'e}sert},
  {Gibson}, {Henry}, {Knutson}, {Lecavelier des Etangs}, {Pont},
  {Vidal-Madjar}, {Williamson}, \& {Wilson}}]{sin15}
{Sing}, D.~K., {Wakeford}, H.~R., {Showman}, A.~P., {Nikolov}, N., {Fortney},
  J.~J., {Burrows}, A.~S., {Ballester}, G.~E., {Deming}, D., {Aigrain}, S.,
  {D{\'e}sert}, J.-M., {Gibson}, N.~P., {Henry}, G.~W., {Knutson}, H.,
  {Lecavelier des Etangs}, A., {Pont}, F., {Vidal-Madjar}, A., {Williamson},
  M.~W., \& {Wilson}, P.~A. 2015, \mnras, 446, 2428

\bibitem[{{S{\l}owikowska} {et~al.}(2016){S{\l}owikowska}, {Krzeszowski},
  {{\.Z}ejmo}, {Reig}, \& {Steele}}]{slo16}
{S{\l}owikowska}, A., {Krzeszowski}, K., {{\.Z}ejmo}, M., {Reig}, P., \&
  {Steele}, I. 2016, \mnras, 458, 759

\bibitem[{{Steele} {et~al.}(2018){Steele}, {Copperwheat}, {Jermak}, {Kennedy},
  \& {Lamb}}]{ste18}
{Steele}, I.~A., {Copperwheat}, C.~M., {Jermak}, H.~E., {Kennedy}, G.~M., \&
  {Lamb}, G.~P. 2018, \mnras, 473, L26

\bibitem[{{Stetson}(1987)}]{ste87}
{Stetson}, P.~B. 1987, \pasp, 99, 191

\bibitem[{{Thompson} {et~al.}(2016){Thompson}, {Scicluna}, {Kemper}, {Geach},
  {Dunham}, {Morata}, {Ertel}, {Ho}, {Dempsey}, {Coulson}, {Petitpas}, \&
  {Kristensen}}]{tho16}
{Thompson}, M.~A., {Scicluna}, P., {Kemper}, F., {Geach}, J.~E., {Dunham},
  M.~M., {Morata}, O., {Ertel}, S., {Ho}, P.~T.~P., {Dempsey}, J., {Coulson},
  I., {Petitpas}, G., \& {Kristensen}, L.~E. 2016, \mnras, 458, L39

\bibitem[{{Vogt} {et~al.}(1994){Vogt}, {Allen}, {Bigelow}, {Bresee}, {Brown},
  {Cantrall}, {Conrad}, {Couture}, {Delaney}, {Epps}, {Hilyard}, {Hilyard},
  {Horn}, {Jern}, {Kanto}, {Keane}, {Kibrick}, {Lewis}, {Osborne},
  {Pardeilhan}, {Pfister}, {Ricketts}, {Robinson}, {Stover}, {Tucker}, {Ward},
  \& {Wei}}]{Vogt1994}
{Vogt}, S.~S., {Allen}, S.~L., {Bigelow}, B.~C., {Bresee}, L., {Brown}, B.,
  {Cantrall}, T., {Conrad}, A., {Couture}, M., {Delaney}, C., {Epps}, H.~W.,
  {Hilyard}, D., {Hilyard}, D.~F., {Horn}, E., {Jern}, N., {Kanto}, D.,
  {Keane}, M.~J., {Kibrick}, R.~I., {Lewis}, J.~W., {Osborne}, J.,
  {Pardeilhan}, G.~H., {Pfister}, T., {Ricketts}, T., {Robinson}, L.~B.,
  {Stover}, R.~J., {Tucker}, D., {Ward}, J., \& {Wei}, M.~Z. 1994, in
  \procspie, Vol. 2198, Instrumentation in Astronomy VIII, ed. D.~L. {Crawford}
  \& E.~R. {Craine}, 362

\bibitem[{{Wallace} {et~al.}(2011){Wallace}, {Hinkle}, {Livingston}, \&
  {Davis}}]{Wallace2011}
{Wallace}, L., {Hinkle}, K.~H., {Livingston}, W.~C., \& {Davis}, S.~P. 2011,
  \apjs, 195, 6

\bibitem[{{Warren} \& {Brandt}(2008)}]{warren08}
{Warren}, S.~G. \& {Brandt}, R.~E. 2008, Journal of Geophysical Research
  (Atmospheres), 113, D14220

\bibitem[{{Wright} \& {Eastman}(2014)}]{barycenter}
{Wright}, J.~T. \& {Eastman}, J.~D. 2014, \pasp, 126, 838

\bibitem[{{Wright} \& {Sigur\dh sson}(2016)}]{WrightSig2016}
{Wright}, J.~T. \& {Sigur\dh sson}, S. 2016, \apjl, 829, L3

\bibitem[{{Wyatt} {et~al.}(2017){Wyatt}, {van Lieshout}, {Kennedy}, \&
  {Boyajian}}]{wya17}
{Wyatt}, M.~C., {van Lieshout}, R., {Kennedy}, G.~M., \& {Boyajian}, T.~S.
  2017, arXiv:1710.05929

\end{thebibliography}


\appendix

\section{Observations}

\subsection{Photometry}

Regular monitoring of \thisstar\ started in 2016 March with the Las Cumbres Observatory (LCOGT) 0.4~m telescope network, which has identical robotic capabilities to the LCOGT 1~m and 2~m telescope networks \citep{bro13}. The northern hemisphere 0.4~m network currently consists of telescopes at three sites: TFN (Canary Islands, Spain), OGG (Hawaii, USA), and since 2017 November, ELP (Texas, USA).     
The scheduled LCOGT requests consisted of Johnson $B$ and Sloan \rp\ and \ip\ images, taking two exposures per sequence with a cadence of $\sim 30$ min. On JD 2,457,892 (UT 2017 May 18; UT dates are used throughout
this paper), a drop in brightness was claimed as significant at more than one site.  We then increased the priority and cadence of the LCOGT $B$\rp\ip\ sequence, and submitted additional $B$\rp\ip\ sequence requests for coverage on the LCOGT 1~m telescopes in Texas, USA (ELP) and California, USA (SQA) and on the 2~m telescope in Hawaii, USA (OGG2). Data are automatically processed by LCOGT servers using \texttt{BANZAI}\footnote{\url{https://github.com/LCOGT/banzai}}
and transferred to local machines where we perform differential photometry for each telescope and filter image stacks using AstroimageJ \citep{col17}. 

Two hundred and thirty one nightly observations of \thisstar\ were acquired from 15 April 2016 to 21 June 2016 and again from 12 November 2016 to 1 July 2017 with the Tennessee State University Celestron 14-inch (C14) automated imaging telescope (AIT) at Fairborn Observatory. See \citet{sin15} for a brief description of the C14 operation and data analysis.  We saw no evidence for variability in the first observing interval to a limit of a few mmag.  \thisstar\ likewise appeared to be constant in the second interval until 2017 May 18, when the C14 revealed the star had dipped in brightness around a percent.  With the Fairborn and LCOGT data both showing signs of the start of a dip in brightness, an alert for triggered observations was executed.

Observations of \thisstar\ were also made with Calvin College's remotely operated telescope in Rehoboth, NM \citep{mol17}. Images in Sloan $g^{\prime}r^{\prime}i^{\prime}z^{\prime}$ filters were taken on ten nights, on 26 April and then for a string of 9 nights beginning 20 May. We used MaxIm\footnote{\url{http://diffractionlimited.com/product/maxim-dl}} to perform differential photometry.

\thisstar\ was observed with the Joan Or\'{o} robotic 0.8~m telescope (TJO) at the Montsec Astronomical Observatory (Catalonia). The star was regularly monitored from 2017 March 14 and the priority and cadence of the observations was increased once the drop in brightness was detected. Sequences of 5 images in Johnson $R$ filter were then obtained one to three times per night and automatically processed with the TJO reduction pipeline \citep{col06}. Differential photometry was performed using AstroimageJ.

The 0.6~m robotic telescope TRAPPIST-North\footnote{\url{http://www.trappist.uliege.be}} \citep{Gil11, Jeh11} observed \thisstar\ starting 2017 May 20.  Each night of observation consisted of exposures of 13~s and 15~s gathered within an ``$I+z$'' filter and Johnson $V$ filter, respectively.  Data reduction was done automatically and consisted of the calibration (bias, dark, flat field) and alignment of the images, and then of the extraction of the fluxes of selected stars by aperture photometry with the {\texttt{DAOPHOT}} software \citep{ste87}.

Observations at Thacher Observatory, on the campus of The Thacher School in Ojai, CA began in the spring of 2017 in the Johnson $V$ band. A variable number of observations of \thisstar\ were performed nightly depending on the weather and telescope demand. Relative fluxes were obtained with aperture photometry and the ratio between \thisstar\ and the sum of the seven reference stars are averaged to produce one measurement per night using a custom pipeline. 

Johnson-Bessel $BVRI$ images of \thisstar\ were acquired with the NITES telescope \citep{mcc14} on La Palma. The data were reduced in Python with {\texttt{CCDPROC}} \citep{cra15} using a master bias, dark, and flat.   A total of 5, 5, 4, and 2 nonvariable comparison stars were used for each of the $B$, $V$, $R$, and $I$ filters, and aperture photometry was extracted using SEP \citep{ber96, bar16}.  

We obtained 60~s exposure photometry using the SBIG STX-16803 CCD mounted on the 1~m telescope of the Wise Observatory in Israel, on a total of 15 nights between 2017 May 19 and 2017 June 21. We alternated between the $B$, $V$, $r^{\prime}$, $i^{\prime}$, and $z^{\prime}$ filters with $3 \times 3$ binning. The images were bias, dark, and flat-field corrected using \texttt{IRAF}\footnote{IRAF is distributed by
the National Optical Astronomy Observatory, which is
operated by AURA, Inc., under a cooperative agreement
with the NSF.}. Aperture photometry was performed using AstroimageJ.  

Spectrophotometric observations were obtained with the OSIRIS long-slit spectrograph on the 10.5~m Gran Telescopio Canarias (GTC). Between 2017 May 17 and 2017 September 9, thirteen pointings were performed, all with a resolution of $R=1000$. Each pointing consisted of a time series of about 30 min duration, from individual shots with 20~sec exposure time. A nearby star of similar brightness (KIC 8462763, $V=11.86$\,mag) was included in the same slit and used as reference in the photometric analysis. The target's spectra were divided into 5 wavelength ranges, and the fluxes at each of these wavelengths are averages of the 30-min time series. A more detailed description and analysis of the GTC observations will be presented in a forthcoming paper (Deeg et al., in prep.). 

\subsection{Spectroscopy}

\subsubsection{Low-resolution}

Over the six-month period beginning on 2017 May 20,
fourteen optical spectra of \thisstar\ were obtained
with the Kast Double Spectrograph mounted on the 3~m
Shane telescope \citep{miller-stone93} 
at Lick Observatory.  Additional optical spectra were acquired
using the 10~m Keck telescopes on 2017 May 29 with the
DEep Imaging Multi-Object Spectrograph
\citep[DEIMOS;][]{fab03} and again on
2017 June 25 with the Low Resolution Imaging
Spectrometer \citep[LRIS;][]{oke95,roc10}.  
All spectra were taken at or near the parallactic
angle \citep{fil82} to minimize slit
losses caused by atmospheric dispersion. Data were
reduced following standard techniques for CCD
processing and spectrum extraction \citep{sil12} utilizing IRAF routines and custom Python and IDL codes\footnote{\url{https://github.com/ishivvers/TheKastShiv}}.  
Low-order polynomial fits to arc-lamp spectra were
used to calibrate the wavelength scale, and small
adjustments derived from night-sky lines in the target
frames were applied.  Observations of appropriate
spectrophotometric standard stars were used to flux
calibrate the spectra.

\subsubsection{High-resolution}

We analyzed high-resolution optical spectra taken 
prior to (2015 October 31, November 27--29, 
and 2016 August 21; five total)
and during (2017 May 20--22; three total)
the {\it Elsie} event 
with the High Resolution Echelle Spectrometer 
\citep[HIRES;][]{Vogt1994}
on the 10~m Keck-I telescope on Mauna Kea, Hawaii.
All spectra were taken with the 
C2 decker and without the iodine cell 
(slit 14$''$.0 length and 0$''$.861 width), 
giving a typical resolution $R \approx 48,000$ 
with signal-to-noise ratios S/N $\approx$ 100--150 per pixel. 
The spectra cover the following wavelength ranges: 
pre-dip 3643--7990~\AA,
in-dip (2017 May 20--21) 3833--6656~\AA,
and in-dip (2017 May 22) 3101--5987~\AA.



\end{document}